\documentclass[12pt]{article}

\usepackage{graphicx}
\usepackage{float}
\usepackage{cite}
\usepackage{amsfonts}
\usepackage{amssymb}
\usepackage{relsize}
\usepackage[compatibility=false]{caption}
\usepackage{subcaption}
\usepackage{xcolor}

\def\gtwid{\mathrel{\raise.3ex\hbox{$>$\kern-.75em\lower1ex\hbox{$\sim$}}}}
\def\ltwid{\mathrel{\raise.3ex\hbox{$<$\kern-.75em\lower1ex\hbox{$\sim$}}}}
\def\square{\kern1pt\vbox{\hrule height 1.2pt\hbox{\vrule width 1.2pt\hskip 3pt
   \vbox{\vskip 6pt}\hskip 3pt\vrule width 0.6pt}\hrule height 0.6pt}\kern1pt}

\begin{document}

\begin{titlepage}

\begin{flushright}
UFIFT-QG-22-02
\end{flushright}

\vskip 2.5cm

\begin{center}
{\bf Reheating with Effective Potentials}
\end{center}

\vskip 1cm

\begin{center}
S. Katuwal$^{1*}$, S. P. Miao$^{2\star}$ and R. P. Woodard$^{1\dagger}$
\end{center}

\begin{center}
\it{$^{1}$ Department of Physics, University of Florida,\\
Gainesville, FL 32611, UNITED STATES}
\end{center}

\begin{center}
\it{$^{2}$ Department of Physics, National Cheng Kung University, \\
No. 1 University Road, Tainan City 70101, TAIWAN}
\end{center}

\vspace{1cm}

\begin{center}
ABSTRACT
\end{center}
We consider reheating for a charged inflaton which is minimally coupled to
electromagnetism. The evolution of such an inflaton induces a time-dependent
mass for the photon. We show how the massive photon propagator can be 
expressed as a spatial Fourier mode sum involving three different sorts of
mode functions, just like the constant mass case. We develop accurate 
analytic approximations for these mode functions, and use them to approximate
the effective force exerted on the inflaton $0$-mode. This effective force
allows one to simply compute the evolution of the inflaton $0$-mode and to
follow the progress of reheating. 

\begin{flushleft}
\end{flushleft}

\vspace{2cm}

\begin{flushleft}
$^{*}$ e-mail: sanjib.katuwal@ufl.edu \\
$^{\star}$ email: spmiao5@mail.ncku.edu.tw \\
$^{\dagger}$ e-mail: woodard@phys.ufl.edu
\end{flushleft}

\end{titlepage}

\section{Introduction}

Scalar-driven inflation is supported by the slow roll of the inflaton down
its potential. At the end of inflation the inflaton begins oscillating, and
its kinetic energy is transferred to ordinary matter during the process of
reheating. The efficiency of this transfer obviously depends on the way the
inflaton is coupled to ordinary matter. Ema et al. have shown that the most
efficient coupling is that of a charged inflaton to electromagnetism
\cite{Ema:2016dny}. 

What happens is that the evolution of a charged inflaton induces a 
time-dependent photon mass which oscillates around zero during reheating. 
The temporal and longitudinal components of the photon diverge as the
mass goes to zero, which makes reheating very efficient. The process has 
been previously studied by discretizing space, carrying out a finite Fourier 
transform, and then numerically evolving the nonlinear system of the inflaton 
plus electromagnetism \cite{Bezrukov:2020txg}. However, the energy transfer
is broadly distributed over so many modes that there is little point to 
including nonlinear effects in the photon field, provided that its response 
to the inflaton $0$-mode is known to all orders. In that case, one merely 
sums the contribution from each photon mode's wave vector, which can be 
accomplished by varying the inflaton effective potential. The goal of this 
paper is to develop a good analytic approximation for the massive photon 
propagator in a time-dependent inflaton background, and then use it to 
compute the quantum-induced, effective force in the equation for the inflaton 
$0$-mode. In this way reheating can be studied by numerically solving a 
nonlocal equation for the inflaton $0$-mode.

This paper consists of five sections, of which the first is this Introduction.
In section 2 we derive a spatial Fourier mode sum for the massive photon
propagator which is valid when the mass becomes time-dependent. Section 3
develops analytic approximations for the temporal and longitudinal modes,
checking them against explicit numerical analysis for a simple model of 
inflation. In section 4 we discuss how these approximations can be used to 
estimate the quantum-induced effective force which controls the process 
of reheating. Section 5 gives our conclusions.

\section{The Massive Photon Propagator}

The purpose of this section is to generalize the massive photon propagator 
from its known form for a constant mass \cite{Katuwal:2021kry} to the case of
a time-dependent mass. The Lagrangian is,
\begin{eqnarray}
\lefteqn{\mathcal{L} = -\frac14 F_{\mu\nu} F_{\rho\sigma} g^{\mu\rho} g^{\nu\sigma} 
\sqrt{-g} } \nonumber \\
& & \hspace{2cm} - \Bigl( \partial_{\mu} \!-\! i q A_{\mu}\Bigr) \varphi \Bigl(
\partial_{\nu} \!+\! i q A_{\nu} \Bigr) \varphi^* g^{\mu\nu} \sqrt{-g} - 
V(\varphi \varphi^*) \sqrt{-g} \; , \qquad \label{Lagrangian}
\end{eqnarray}
where $\varphi$ is the inflaton and $F_{\mu\nu} \equiv \partial_{\mu} A_{\nu} - 
\partial_{\nu} A_{\mu}$ is the electromagnetic field strength. We work on a 
general homogeneous, isotropic and spatially flat geometry in $D$-dimensional, 
conformal coordinates, with Hubble parameter $H$ and first slow roll parameter 
$\epsilon$,
\begin{equation}
ds^2 = a^2 \Bigl[-d\eta^2 + d\vec{x} \!\cdot\! d\vec{x}\Bigr] \qquad , \qquad
H \equiv \frac{\partial_0 a}{a^2} \quad , \quad \epsilon \equiv -
\frac{\partial_0 H}{a H^2} \; . \label{geometry}
\end{equation}
The section first reviews the constant mass case, and then makes the 
generalizations necessary to incorporate a time-dependent mass.

\subsection{Constant Mass}

When the photon's mass is constant its propagator $i[\mbox{}_{\mu} 
\Delta_{\rho}](x;x')$ is transverse,
\begin{equation}
\partial_{\mu} \Bigl\{ \!\sqrt{-g(x)} \, g^{\mu\nu}(x) \, i\Bigl[\mbox{}_{\nu}
\Delta_{\rho}\Bigr](x;x')\Bigr\} = 0 = \partial'_{\rho} \Bigl\{\! \sqrt{-g(x')} \,
g^{\rho\sigma}(x') \, i\Bigl[\mbox{}_{\mu} \Delta_{\sigma}\Bigr](x;x') \Bigr\} .
\label{transverse}
\end{equation}
Its propagator equation reflects this transversality \cite{Katuwal:2021kry,
Tsamis:2006gj},
\begin{eqnarray}
\lefteqn{ \sqrt{-g} \Bigl[ \square^{\mu\nu} - R^{\mu\nu} - M^2 g^{\mu\nu}\Bigr]
i \Bigl[\mbox{}_{\nu} \Delta_{\rho}\Bigr](x;x') } \nonumber \\
& & \hspace{3.5cm} = \delta^{\mu}_{~\rho} i\delta^D(x \!-\! x') + \sqrt{-g(x)} \,
g^{\mu\nu}(x) \partial_{\nu} \partial'_{\rho} i\Delta(x;x') \; . \qquad
\label{constMprop}
\end{eqnarray}
Here $\square^{\mu\nu}$ is the vector d'Alembertian, $R^{\mu\nu}$ is the Ricci 
tensor and $i\Delta(x;x')$ is the propagator of a massless, minimally coupled 
scalar,
\begin{equation}
\partial_{\mu} \Bigl[\sqrt{-g} \, g^{\mu\nu} \partial_{\nu} i\Delta(x;x') \Bigr]
= i\delta^D(x \!-\! x') \; . \label{MMCSprop}
\end{equation}

The solution to (\ref{transverse}-\ref{constMprop}) can be expressed as a spatial
Fourier mode sum over three sorts of polarizations \cite{Katuwal:2021kry},
\begin{eqnarray}
\lefteqn{ i\Bigl[\mbox{}_{\mu} \Delta_{\rho}\Bigr](x;x') = \int \!\! 
\frac{d^{D-1}k}{(2\pi)^{D-1}} \sum_{\lambda = t,u,v} s_{\lambda} \Biggl\{ 
\theta(\Delta \eta) \mathcal{A}_{\mu}(x;\vec{k},\lambda) \mathcal{A}_{\nu}^*(x';
\vec{k},\lambda) } \nonumber \\
& & \hspace{6cm} + \theta(-\Delta \eta) \mathcal{A}_{\mu}^*(x;\vec{k},\lambda)
\mathcal{A}_{\nu}(x';\vec{k},\lambda) \Biggr\} , \qquad \label{constMmodesum}
\end{eqnarray}
where $\Delta \eta \equiv \eta - \eta'$. Longitudinal photons correspond to
$\lambda = t$ and have $s_t = -1$ with,
\begin{equation}
\mathcal{A}_{\mu}(x;\vec{k},t) = \frac{\partial_{\mu}}{M} \Bigl[ t(\eta,k)
e^{i\vec{k} \cdot \vec{x}} \Bigr] \;\; , \;\; \Bigl[ \mathcal{D} \partial_0 +
k^2\Bigr] t = 0 \;\; , \;\; t \cdot \partial_0 t^* - \partial_0 t \cdot t^* = 
\frac{i}{a^{D-2}} , \label{tmodes}
\end{equation}
where $\mathcal{D} \equiv \partial_0 + (D-2) a H$. Temporal photons correspond
to $\lambda = u$ and have $s_u = +1$ with,
\begin{eqnarray}
\mathcal{A}_{\mu}(x;\vec{k},u) = \frac{\overline{\partial}_{\mu}}{M} \Bigl[ 
u(\eta,k) e^{i\vec{k} \cdot \vec{x}} \Bigr] & , & \overline{\partial}_0 \equiv k
\;\; , \;\; \overline{\partial}_m \equiv \frac{-i k_m}{k} \mathcal{D} \; , 
\qquad \\
\Bigl[ \partial_0 \mathcal{D} + k^2 + a^2 M^2\Bigr] u = 0 & , & u 
\cdot \partial_0 u^* - \partial_0 u \cdot u^* = \frac{i}{a^{D-2}} . \qquad
\label{umodes}
\end{eqnarray}
Transverse spatial photons correspond to $\lambda = v$ and have $s_v = +1$ with,
\begin{eqnarray}
\mathcal{A}_{\mu}(x;\vec{k},v) = \epsilon_{\mu}(\vec{k},v) \, v(\eta,k) 
e^{i\vec{k} \cdot \vec{x}} & , & \epsilon_0 = 0 \;\; , \;\; k_m \epsilon_m = 0 
\; , \qquad \label{vmodesA} \\
\Bigl[ \partial^2_0 + (D \!-\! 4) a H \partial_0 + k^2 + a^2 M^2\Bigr] v = 0 & , & 
v \cdot \partial_0 v^* - \partial_0 v \cdot v^* = \frac{i}{a^{D-4}} , \qquad
\label{vmodesB}
\end{eqnarray}
where the sum over the $(D-2)$ spatial polarizations gives,
\begin{equation}
\sum_{v} \epsilon_i(\vec{k},v) \times \epsilon^*_j(\vec{k},v) = \delta_{ij} - 
\frac{k_i k_j}{k^2} \; . \label{polsum}
\end{equation}

\subsection{Time-Dependent Mass}

To understand the case of a time-dependent mass we must consider the vector
and scalar field equations,
\begin{eqnarray}
\lefteqn{ \frac{\delta S}{\delta A_{\mu}} = \partial_{\nu} \Bigl[ \sqrt{-g} \,
g^{\nu\rho} g^{\mu\sigma} F_{\rho\sigma} \Bigr] } \nonumber \\
& & \hspace{2.5cm} + iq \Bigl[ \varphi \!\cdot\! \Bigl( \partial_{\nu} \!+\! i q
A_{\nu}\Bigr) \varphi^* - \Bigl( \partial_{\nu} \!-\! i q A_{\nu}\Bigr) \varphi
\!\cdot\! \varphi^* \Bigr] g^{\mu\nu} \sqrt{-g} \; , \qquad \label{vector} \\
\lefteqn{ \frac{\delta S}{\delta \varphi^*} = \Bigl(\partial_{\mu} \!-\! i q
A_{\mu}\Bigr) \Bigl[ \sqrt{-g} \, g^{\mu\nu} \Bigl(\partial_{\nu} \!-\! i q
A_{\nu}\Bigr) \varphi \Bigr] - \varphi V'(\varphi \varphi^*) \sqrt{-g} \; .} 
\label{scalar}
\end{eqnarray}
The $0$-th order inflaton is $\varphi_0(\eta)$ which is real and obeys 
the equation,
\begin{equation}
\partial_0 \Bigl[ a^{D-2} \partial_0 \varphi_0\Bigr] + a^{D} \varphi_0
V'(\varphi_0^2) = 0 \; . \label{0order}
\end{equation}
The first order perturbations are $A_{\mu}(x)$ and the real fields 
$\alpha(x)$ and $\beta(x)$,
\begin{equation}
\varphi(x) = \varphi_0(\eta) + \alpha(x) + i \beta(x) \; .
\end{equation}
The first order contribution to the vector equation (\ref{vector}) is,
\begin{equation}
\partial_{\nu} \Bigl[ \sqrt{-g} \, g^{\nu\rho} g^{\mu\sigma} F_{\rho\sigma}
\Bigr] - 2 q^2 \varphi_0^2 \Bigl[ A_{\nu} - \partial_{\nu} \Bigl( 
\frac{\beta}{q \varphi_0}\Bigr) \Bigr] \sqrt{-g} \, g^{\nu\mu} = 0 \; . 
\label{vector1}
\end{equation}
The photon mass is $M^2 \equiv 2 q^2 \varphi_0^2$. Note from equation 
(\ref{vector1}) that antisymmetry of the field strength tensor implies,
\begin{equation} 
\partial_{\mu} \Bigl[ M^2 \sqrt{-g} \, g^{\mu\nu} \Bigl(A_{\nu} - 
\partial_{\nu} \Bigl( \frac{\beta}{q \varphi_0} \Bigr) \Bigr] = 0 \; . 
\label{betaeqn}
\end{equation}
This constraint is identical to the imaginary part of the first order
contribution to the scalar equation (\ref{scalar}). The analogous real 
part is,
\begin{equation}
\partial_{\mu} \Bigl[ \sqrt{-g} \, g^{\mu\nu} \partial_{\nu} \alpha\Bigr] 
- \sqrt{-g} \Bigl[ V'(\varphi_0^2) + 2 \varphi_0^2 V''(\varphi_0^2)\Bigr] 
\alpha = 0 \; . \label{alphaeqn}
\end{equation}

Relations (\ref{vector1}) and (\ref{betaeqn}) demonstrate that the Higgs 
mechanism continues to function when the scalar background $\varphi_0$ 
depends upon spacetime. To simplify the subsequent analysis, we will
absorb (``eat'') the imaginary part of the scalar perturbation into the 
vector field as usual,
\begin{equation}
A_{\mu} - \partial_{\mu} \Bigl( \frac{\beta}{q \varphi_0} \Bigr) 
\longrightarrow A_{\mu} \; . \label{eating}
\end{equation}
We can also use the conformal coordinate relation $g_{\mu\nu} = a^2 
\eta_{\mu\nu}$ to provide simple expressions for (\ref{vector1}) and
(\ref{betaeqn}),
\begin{equation}
\partial_{\nu} \Bigl[ a^{D-4} F^{\nu\mu} \Bigr] - M^2 a^{D-2} A^{\mu} 
= 0 \qquad \Longrightarrow \qquad \partial_{\mu} \Bigl[ M^2 a^{D-2} 
A^{\mu} \Bigr] = 0 \; , \label{vector2} 
\end{equation}
where $F^{\nu\mu} \equiv \eta^{\nu\rho} \eta^{\mu\sigma} F_{\rho\sigma}$
and $A^{\mu} \equiv \eta^{\mu\nu} A_{\nu}$. The $3+1$ decomposition of
the constraint on the right hand side of (\ref{vector2}) is,
\begin{equation}
\Bigl[ \mathcal{D} + \frac{2 \partial_0 M}{M}\Bigr] A_0 - 
\partial_m A_m = 0 \qquad , \qquad \mathcal{D} \equiv \partial_0 + 
(D\!-\!2) a H \; . \label{constraint}
\end{equation}
Relation (\ref{constraint}) permits us to $3+1$ decompose the left hand 
side of (\ref{vector2}) to,
\begin{eqnarray}
\Bigl[ \partial_0 \Bigl( \mathcal{D} + \frac{2 \partial_0 M}{M} \Bigr)
- \nabla^2 + a^2 M^2\Bigr] A_0 &\!\!\! = \!\!\!& 0 \; , 
\label{vector3A} \\
2 \Bigl( a H + \frac{\partial_0 M}{M}\Bigr) \partial_m A_0 + \Bigl[
\partial_0^2 + (D\!-\!4) a H \partial_0 - \nabla^2 + a^2 M^2\Bigr] A_m
&\!\!\! = \!\!\!& 0 \; . \label{vector3B}
\end{eqnarray}

Equations (\ref{constraint}-\ref{vector3B}) are satisfied by three 
polarizations of spatial plane waves whose associated mode functions
are $t(\eta,k)$, $u(\eta,k)$ and $v(\eta,k)$. Our notation is that a
``tilde'' over a differential operator such as $\partial_0$ or $\mathcal{D}$
indicates the addition of $\partial_0 M/M$, whereas a ``hat'' denotes
subtraction of the same quantity,
\begin{equation}
\widetilde{\mathcal{D}} \equiv \mathcal{D} + \frac{\partial_0 M}{M} 
\qquad , \qquad \widehat{\partial}_0 \equiv \partial_0 - 
\frac{\partial_0 M}{M} \; . \label{notation}
\end{equation}
What we term {\it Longitudinal photons} have the form,
\begin{equation}
\mathcal{A}_0(x;\vec{k},t) = \frac{\widehat{\partial}_0 t(\eta,k)}{M(\eta)} \,
e^{i \vec{k} \cdot \vec{x}} \qquad , \qquad \mathcal{A}_m(x;\vec{k},t) =
\frac{i k_m t(\eta,k)}{M(\eta)} \, e^{i \vec{k} \cdot \vec{x}} \; ,
\label{tAs}
\end{equation}
where the mode function $t(\eta,k)$ obeys,\footnote{Although 
$\mathcal{A}_{\mu}(x;\vec{k},t)$ satisfies (\ref{constraint}), it does not 
quite obey equations (\ref{vector3A}-\ref{vector3B}), but rather the relation
$\partial_{\nu} [a^{D-4} \mathcal{F}^{\nu\mu}(x;\vec{k},t)] = 0$.}
\begin{equation}
\Bigl[ \widetilde{\mathcal{D}} \widehat{\partial}_0 + k^2\Bigr] t = 0 \qquad , 
\qquad t \cdot \partial_0 t^* - \partial_0 t \cdot t^* = \frac{i}{a^{D-2}} \; .
\label{teqn}
\end{equation}
{\it Temporal photons} take the form,
\begin{equation}
\mathcal{A}_0(x;\vec{k},u) = \frac{k u(\eta,k)}{M(\eta)} \, e^{i \vec{k} \cdot \vec{x}} 
\qquad , \qquad \mathcal{A}_m(x;\vec{k},u) = - \frac{i k_m \widetilde{\mathcal{D}} 
u(\eta,k)}{k M(\eta)} \, e^{i \vec{k} \cdot \vec{x}} \; , \label{uAs}
\end{equation}
where the mode function $u(\eta,k)$ obeys,
\begin{equation}
\Bigl[ \widehat{\partial}_0 \widetilde{\mathcal{D}} + k^2 + a^2 M^2 \Bigr] u = 0 
\qquad , \qquad u \cdot \partial_0 u^* - \partial_0 u \cdot u^* = \frac{i}{a^{D-2}} 
\; . \label{ueqn}
\end{equation}
The tendency for longitudinal and temporal photons to diverge when the mass 
$M(\eta)$ passes through zero is obvious from expressions (\ref{tAs}) and 
(\ref{uAs}). In contrast, the time-dependent mass makes no change at all in 
relations (\ref{vmodesA}-\ref{polsum}) for the {\it Transverse spatial photons}, 
and these polarizations remain finite as the mass passes through zero.

A time-dependent mass makes no change in mode sum (\ref{constMmodesum}) for 
the propagator. However, the propagator obeys a revised version of the
constraint equation (\ref{transverse}),
\begin{equation}
\partial^{\mu} \Bigl\{a^{D-2} M^2 i \Bigl[\mbox{}_{\mu} \Delta_{\rho}\Bigr](x;x')
\Bigr\} = 0 = \partial^{\prime \rho} \Bigl\{ {a'}^{D-2} {M'}^2 i\Bigl[ \mbox{}_{\mu}
\Delta_{\rho}\Bigr](x;x') \Bigr\} \; . \label{newconstraint}
\end{equation}
The propagator equations analogous to (\ref{constMprop}-\ref{MMCSprop}) can be
given in terms of the massive photon kinetic operator,
\begin{equation}
\mathcal{D}^{\mu\nu} \equiv \partial_{\alpha} \Bigl[ a^{D-4} \Bigl( \eta^{\mu\nu}
\partial^{\alpha} - \eta^{\alpha\nu} \partial^{\mu}\Bigr) \Bigr] - a^{D-2} M^2
\eta^{\mu\nu} \; . \label{kineticop}
\end{equation}
The revised versions of (\ref{constMprop}-\ref{MMCSprop}) are,
\begin{eqnarray}
\mathcal{D}^{\mu\nu} i\Bigl[\mbox{}_{\nu} \Delta_{\rho}\Bigr](x;x') & \!\!\! =
\!\!\!& \delta^{\mu}_{~\rho} i\delta^D(x \!-\! x') + \frac{a^{D-2} M}{M'}
\widehat{\partial}^{\mu} \widehat{\partial}'_{\rho} i \Delta_{t}(x;x') \; ,
\qquad \label{newprop} \\
\frac1{M} \partial^{\mu} \Bigl[ a^{D-2} M \widehat{\partial}_{\mu} 
i\Delta_{t}(x;x') \Bigr] &\!\!\! = \!\!\!& i\delta^D(x \!-\! x') \; . \qquad
\label{newtprop}
\end{eqnarray}

\section{Approximating the Amplitudes}

The purpose of this section is to develop analytic approximations for the
crucial mode functions $t(\eta,k)$ and $u(\eta,k)$. We begin by giving a
dimensionless formulation of the problem. This formalism is then employed
to derive good analytic approximations for first, the longitudinal amplitude
and then, the temporal amplitude. At each stage these approximations are
checked against explicit numerical evolution in a simple mode of inflation.

\subsection{Dimensionless Formulation}

It is best to change the evolution variable from conformal time $\eta$ to
the number of e-foldings from the start of inflation, $n \equiv \ln[a(\eta)]$,
\begin{equation}
\partial_0 = a H \frac{\partial}{\partial n} \qquad , \qquad \partial_0^2
= a^2 H^2 \Bigl[ \frac{\partial^2}{\partial n^2} + (1 \!-\! \epsilon)
\frac{\partial}{\partial n}\Bigr] \; .
\end{equation}
We can also use factors of $8 \pi G$ to make the inflaton, the Hubble parameter
and the scalar potential dimensionless,
\begin{equation}
\psi(n) \equiv \sqrt{8\pi G} \, \varphi_0(\eta) \quad , \quad
\chi(n) \equiv \sqrt{8\pi G} \, H(\eta) \quad , \quad
U(\psi^2) \equiv (8\pi G)^2 V(\varphi^2_0) \; . \label{dimgeom}
\end{equation}
This gives dimensionless forms for the classical Friedmann equations, and for
the inflaton evolution equation, 
\begin{eqnarray}
\frac12 (D \!-\! 2) (D \!-\! 1) \chi^2 &\!\!\! = \!\!\!& \chi^2 {\psi'}^2 + 
U(\psi^2) \; , \qquad \label{Friedmann1} \\
-\frac12 (D\!-\!2) \Bigl[ (D\!-\! 1) - 2 \epsilon\Bigr] \chi^2 &\!\!\! = 
\!\!\!& \chi^2 {\psi'}^2 - U(\psi^2) \; , \qquad \label{Friedmann2} \\
0 &\!\!\! = \!\!\!& \chi^2 \Bigl[ \psi'' + (D\!-\!1\!-\!\epsilon) \psi'\Bigr] 
+ \psi U'(\psi^2) \; . \qquad \label{inflatoneqn}
\end{eqnarray}

Factors of $8\pi G$ can be extracted to give similar dimensionless forms for
the time-dependent mass $M^2(\eta) \equiv 2 q^2 \varphi_0^2(\eta)$ and the wave
number $k^2$,
\begin{equation}
\mu^2(n) \equiv 8\pi G M^2(\eta) = 2 q^2 \psi^2(n) \qquad , \qquad \kappa^2
\equiv 8 \pi G k^2 \; . \label{dimparams}
\end{equation}
We define the dimensionless Longitudinal and Temporal amplitudes as,
\begin{equation}
\mathcal{T}(n,\kappa) \equiv \ln\Bigl[ \frac{\vert t(\eta,k)\vert^2}{
\sqrt{8\pi G}}\Bigr] \qquad , \qquad \mathcal{U}(n,\kappa) \equiv
\ln\Bigl[\frac{\vert u(\eta,k)\vert^2}{\sqrt{8\pi G}}\Bigr] \; . \label{Amps}
\end{equation}
By combining the mode equations and Wronskians (\ref{teqn}) and (\ref{ueqn}) 
for each mode we can infer a single nonlinear relation for the associated
amplitudes \cite{Romania:2011ez,Romania:2012tb,Brooker:2015iya},
\begin{eqnarray}
\mathcal{T}'' + \frac12 {\mathcal{T}'}^2 + (D \!-\! 1 \!-\! \epsilon) 
\mathcal{T}' + \frac{2 \kappa^2 e^{-2n}}{\chi^2} + \frac{2 \mu_t^2}{\chi^2}
- \frac{e^{-2 [\mathcal{T} + (D-1) n]}}{2 \chi^2} &\!\!\! = \!\!\! & 0 
\; , \qquad \label{Teqn} \\
\mathcal{U}'' + \frac12 {\mathcal{U}'}^2 + (D \!-\! 1 \!-\! \epsilon) 
\mathcal{U}' + \frac{2 \kappa^2 e^{-2n}}{\chi^2} + \frac{2 \mu_u^2}{\chi^2}
- \frac{e^{-2 [\mathcal{U} + (D-1) n]}}{2 \chi^2} &\!\!\! = \!\!\! & 0 
\; , \qquad \label{Ueqn}
\end{eqnarray}
where a prime denotes differentiation with respect to $n$ and the two 
masses are,
\begin{eqnarray}
\frac{\mu^2_t}{\chi^2} &\!\!\! \equiv \!\!\!& -(D\!-\!1\!-\!\epsilon) 
\frac{\mu'}{\mu} - \frac{\mu''}{\mu} \; , \qquad \label{tmass} \\
\frac{\mu^2_u}{\chi^2} &\!\!\! \equiv \!\!\!& (D\!-\!2) (1\!-\! \epsilon)
+ \frac{\mu^2}{\chi^2} - (D\!-\!3\!+\!\epsilon) \frac{\mu'}{\mu} + 
\Bigl( \frac{\mu'}{\mu}\Bigr)' - \Bigl( \frac{\mu'}{\mu}\Bigr)^2 \; . 
\qquad \label{umass}
\end{eqnarray}
Because $\mu^2(n) = 2 q^2 \psi^2(n)$ we can use the inflaton $0$-mode
equation (\ref{inflatoneqn}) to simplify the $t$-mode mass,
\begin{equation}
\frac{\mu^2_{t}}{\chi^2} = -\frac{[\psi'' + (D\!-\!1\!-\!\epsilon) \psi']}{
\psi} = \frac{U'(\psi^2)}{\chi^2} \; . \label{tmasssimp}
\end{equation}

In order to follow the amplitudes numerically one must use a specific 
model of inflation. For simplicity we have chosen the quadratic mass
model, $U = c^2 \psi^2$, even though its prediction for the 
tensor-to-scalar ratio is disfavored by the data \cite{Planck:2018vyg,
Tristram:2020wbi}. The Slow Roll Approximation gives analytic expressions 
for this model which are accurate until almost the end of inflation,
\begin{equation}
\psi(n) \simeq \sqrt{\psi_0^2 \!-\! 2n} \quad , \quad \chi(n) \simeq
\frac{c}{\sqrt{3}} \sqrt{\psi_0^2 \!-\! 2n} \quad ,\quad \epsilon(n)
\simeq \frac1{\psi_0^2 \!-\! 2n} \; , \label{slowroll}
\end{equation}
where $\psi_0$ is the initial value of the dimensionless inflaton 
$0$-mode. About 56 e-foldings of inflation results from the choice 
$\psi_0 = 10.6$. To estimate the constant $c$, note that modes which 
experience 1st horizon crossing at e-folding $n_1$ (that is, $\kappa = 
\chi(n_1) e^{n_1}$) have the following approximate scalar power spectrum
and spectral index,
\begin{equation}
\Delta^2_{\mathcal{R}}(n_1) \simeq \frac1{8\pi^2} \frac{\chi^2(n_1)}{\epsilon(n_1)}
\qquad \Longrightarrow \qquad 1 - n_s \simeq 2 \epsilon +
\frac{\epsilon'}{\epsilon} \; . \label{CMB}
\end{equation}
Hence the observed scalar spectral index is consistent with $\psi_0 =
10.6$, and the observed scalar amplitude with the choice of $c = 7.1 
\times 10^{-6}$ \cite{Planck:2018vyg,Tristram:2020wbi}. We must also
choose a specific value for the charge $q$. Using $q^2 = 1/137$ would cause
the classical potential of $U = c^2 \psi^2$ to be completely overwhelmed
by the 1-loop Coleman-Weinberg correction of $\Delta U \simeq 3/64\pi^2 
\times \mu^4 \ln(\mu^2/s^2)$, where $s$ is the dimensionless 
renormalization scale \cite{Miao:2015oba}. Choosing the much smaller 
value of $q = 1.2 \times 10^{-6}$ reduces the 1-loop correction to a 
negligible tenth of a percent effect at the start of inflation.

\begin{figure}[H]
\centering
\includegraphics[width=4.3cm]{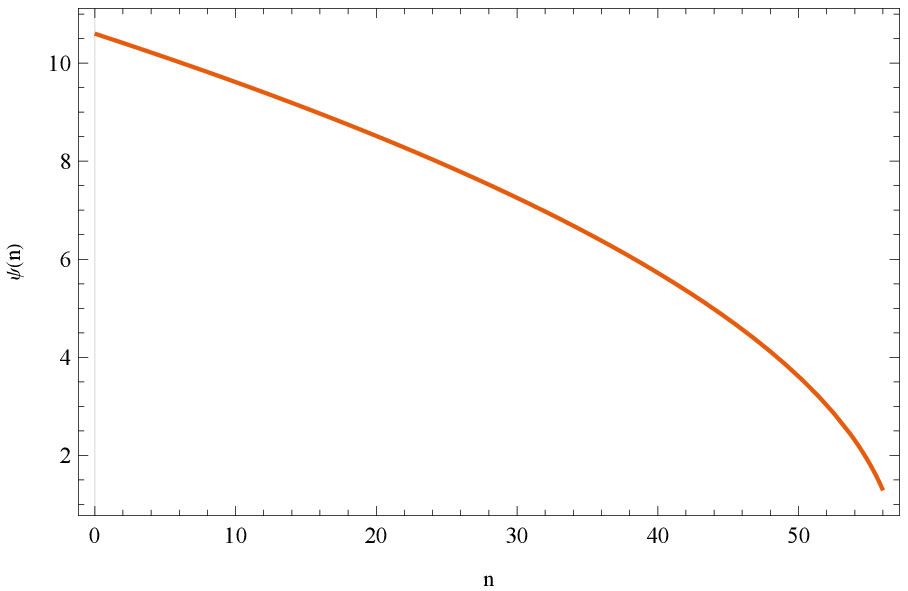}
\includegraphics[width=4.3cm]{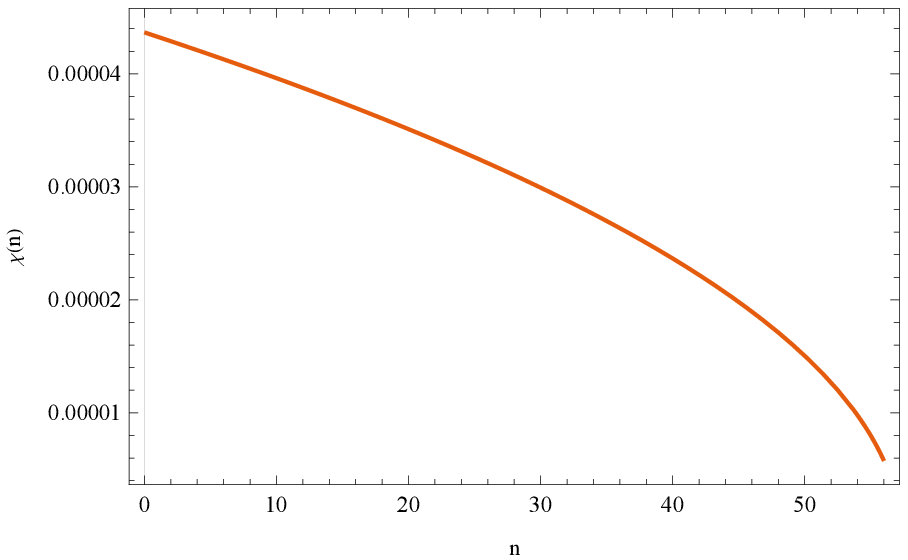}
\includegraphics[width=4.3cm]{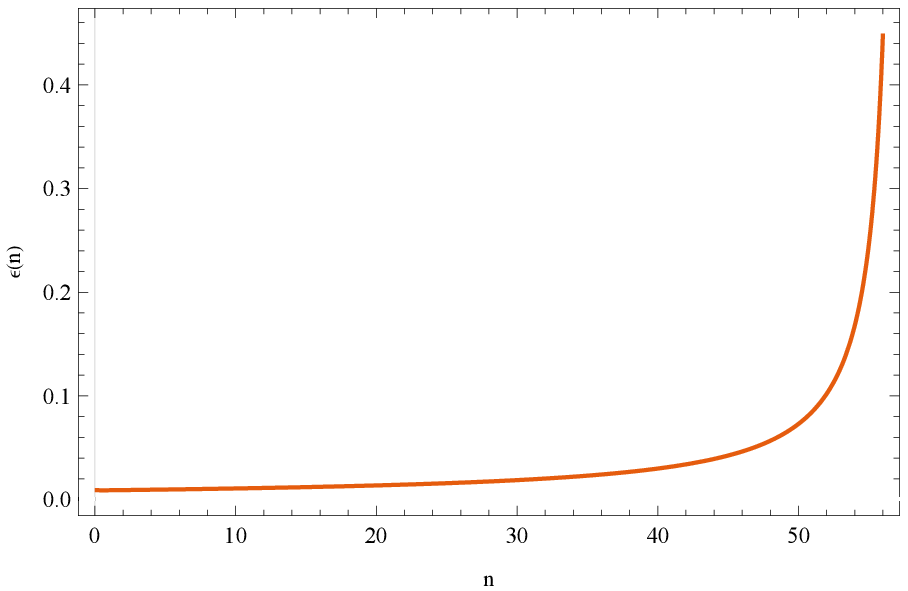}
\caption{\footnotesize Plots of $\psi(n)$ --- on the left --- $\chi(n)$ ---
in the center --- and $\epsilon(n)$ --- on the right --- for $0 \leq n \leq 
56$.}
\label{Earlygeom}
\end{figure}

\begin{figure}[H]
\centering
\includegraphics[width=4.3cm]{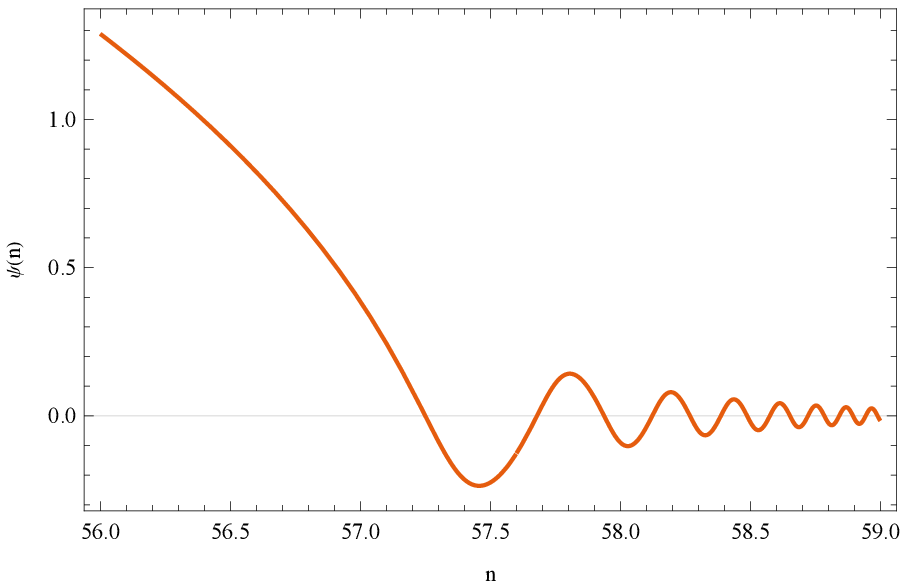}
\includegraphics[width=4.3cm]{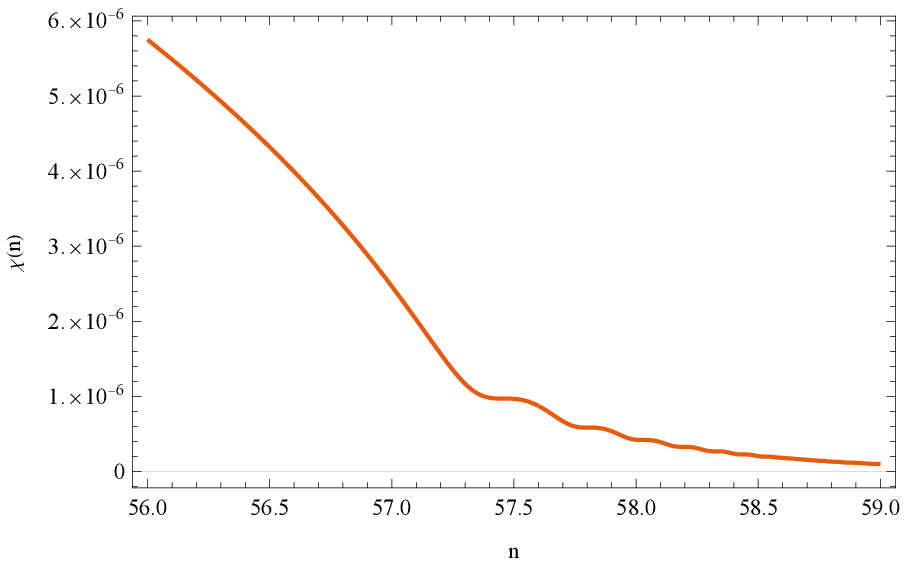}
\includegraphics[width=4.3cm]{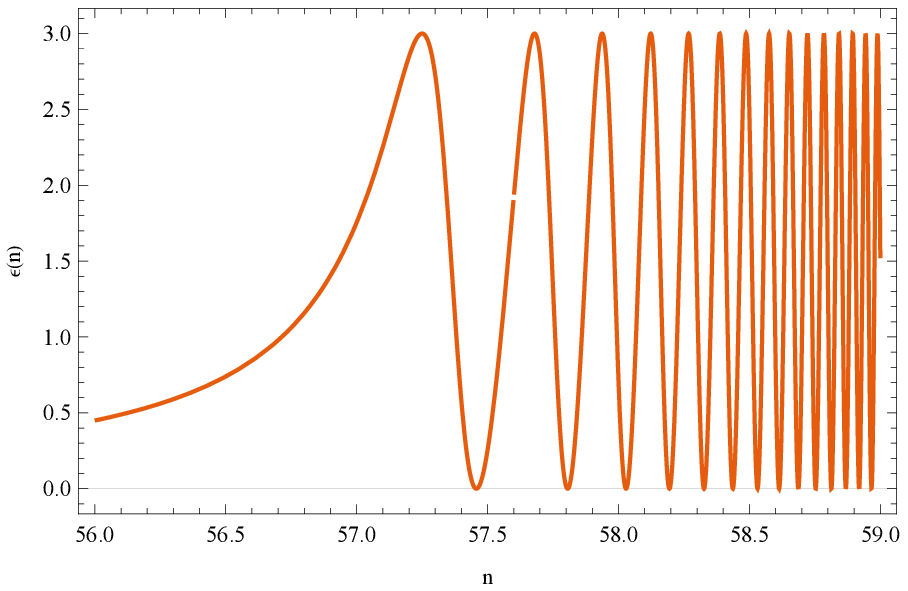}
\caption{\footnotesize Plots of $\psi(n)$ --- on the left --- $\chi(n)$ ---
in the center --- and $\epsilon(n)$ --- on the right --- for $56 \leq n 
\leq 59$.}
\label{Lategeom}
\end{figure}

Once we have a specific model it is possible to understand the magnitudes
of the various terms. Figure~\ref{Earlygeom} shows the dimensionless scalar, 
the dimensionless Hubble parameter and the first slow roll parameter while
inflation is occurring ($\epsilon < 1$). The slow roll approximations 
(\ref{slowroll}) are excellent during this period.
\noindent Figure~\ref{Lategeom} shows the same three quantities through the
end of inflation (which occurs at $n_e \simeq 56.7$) under the assumption
that the classical relations (\ref{Friedmann1}-\ref{inflatoneqn}) are not
corrected by the quantum effects we seek to incorporate. During this phase 
the inflaton oscillates around $\psi = 0$ with decreasing amplitude and
increasing frequency, while the first slow roll parameter oscillates in the 
range $0 \leq \epsilon \leq 3$. Because $\epsilon = {\psi'}^2$, the first 
slow roll parameter vanishes at extrema of $\psi(n)$, and it reaches its 
maximum (of $\epsilon(n) = 3$) when $\psi(n) = 0$. Of course the dimensionless 
Hubble parameter is monotonically decreasing; this decrease is rapid when 
$\epsilon \simeq 3$, and slow when $\epsilon \simeq 0$. 

\begin{figure}[H]
\centering
\includegraphics[width=4.3cm]{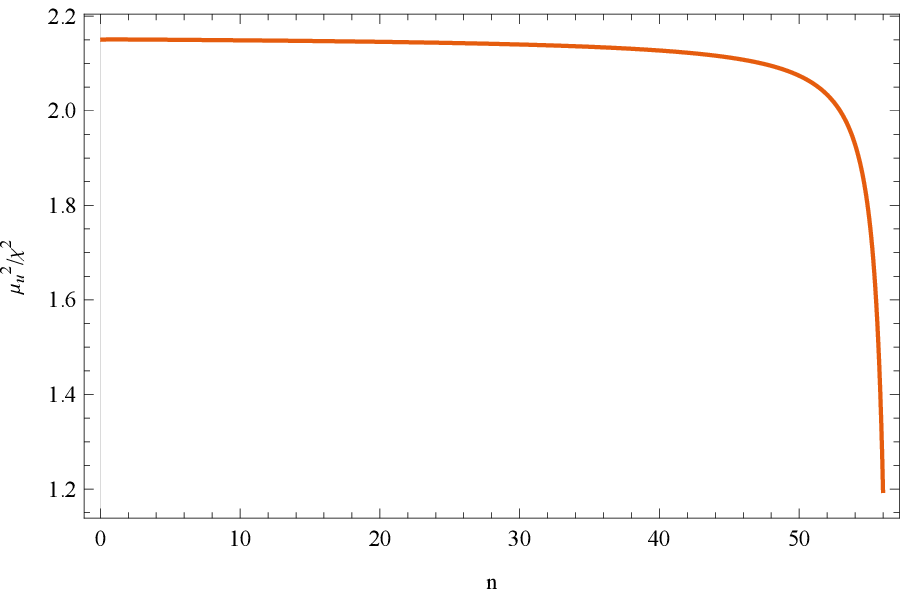}
\includegraphics[width=4.3cm]{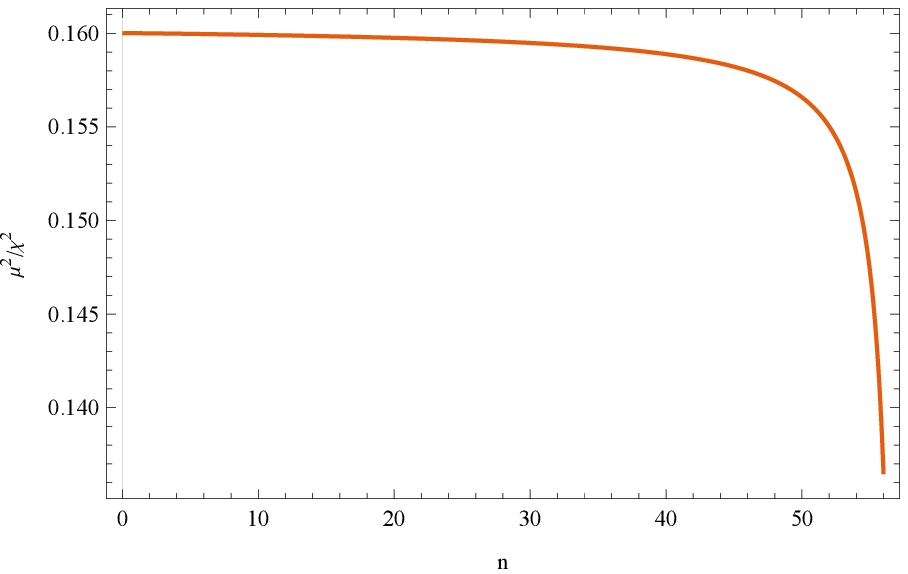}
\includegraphics[width=4.3cm]{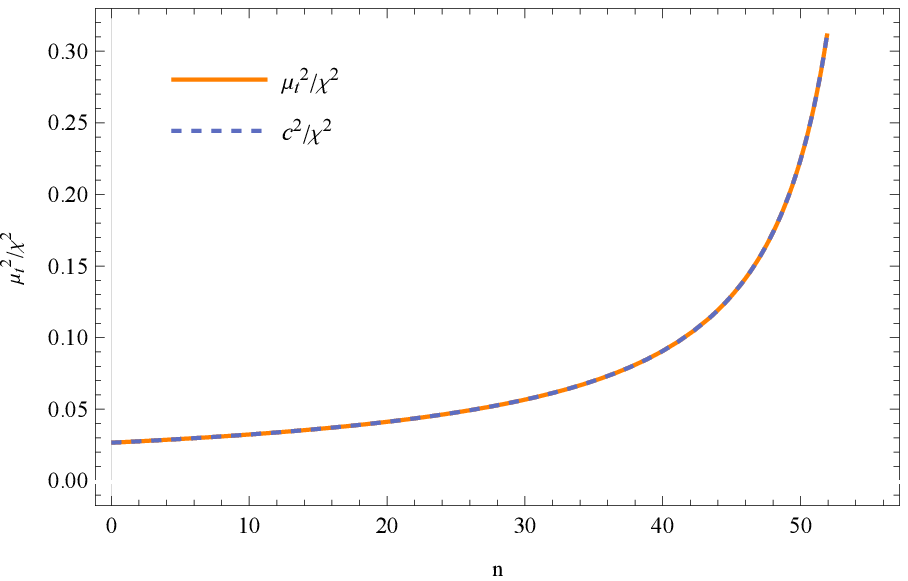}
\caption{\footnotesize Plots of $\mu_u^2(n)/\chi^2(n)$ --- on the left --- 
$\mu^2(n)/\chi^2(n)$ --- in the center --- and $\mu^2_t(n)/\chi^2(n)$ --- 
on the right --- for $0 \leq n \leq 56$.}
\label{Earlymass}
\end{figure}

\begin{figure}[H]
\centering
\includegraphics[width=4.3cm]{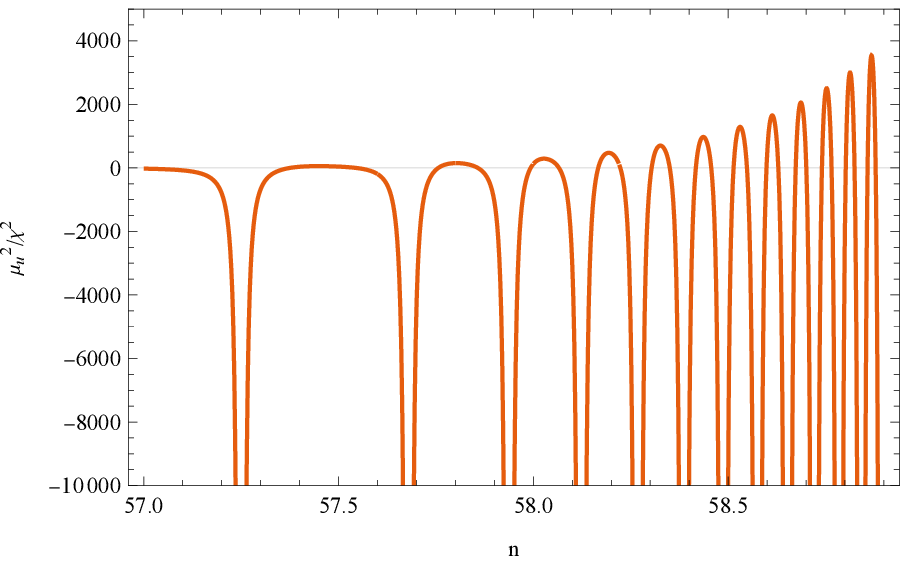}
\includegraphics[width=4.3cm]{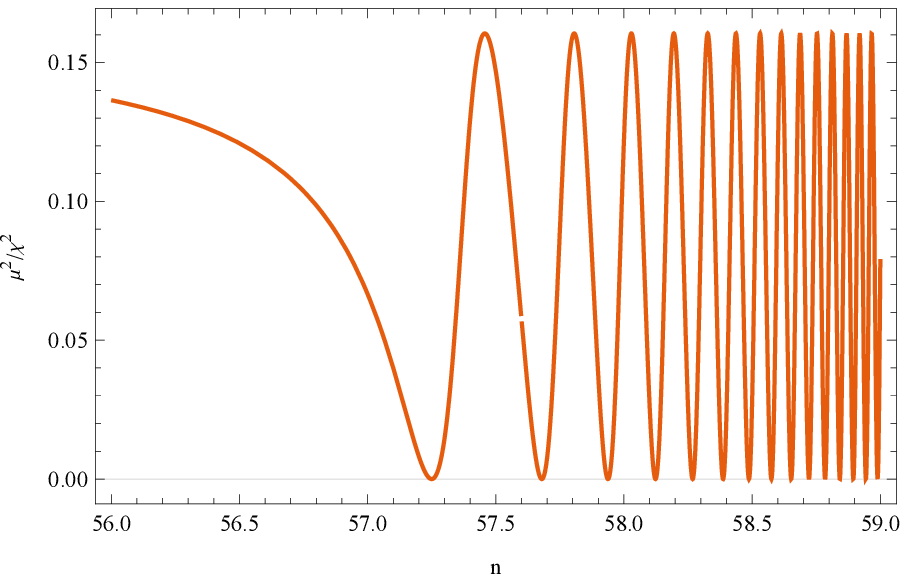}
\includegraphics[width=4.3cm]{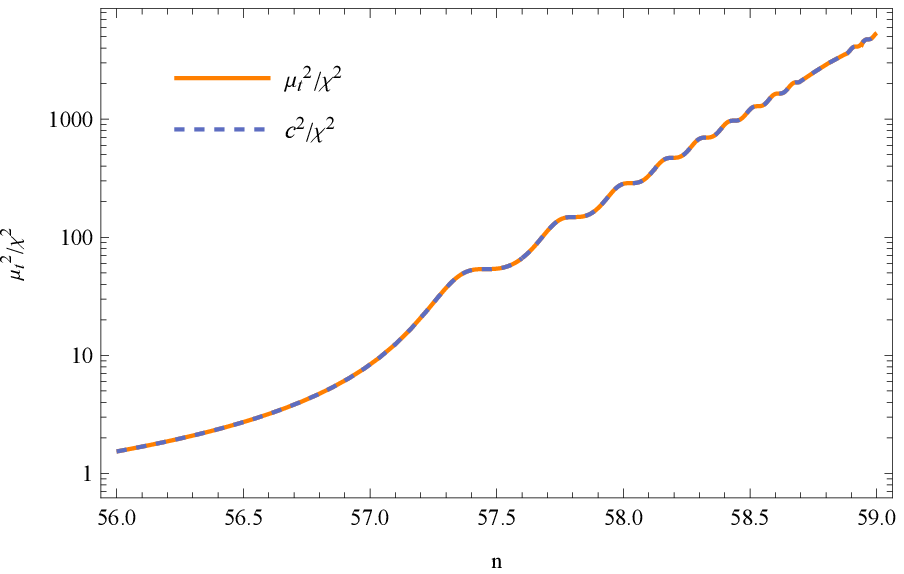}
\caption{\footnotesize Plots of $\mu_u^2(n)/\chi^2(n)$ --- on the left --- 
$\mu^2(n)/\chi^2(n)$ --- in the center --- and $\mu^2_t(n)/\chi^2(n)$ --- 
on the right --- for $56 \leq n \leq 59$.}
\label{Latemass}
\end{figure}

\begin{figure}[H]
\centering
\includegraphics[width=6.5cm]{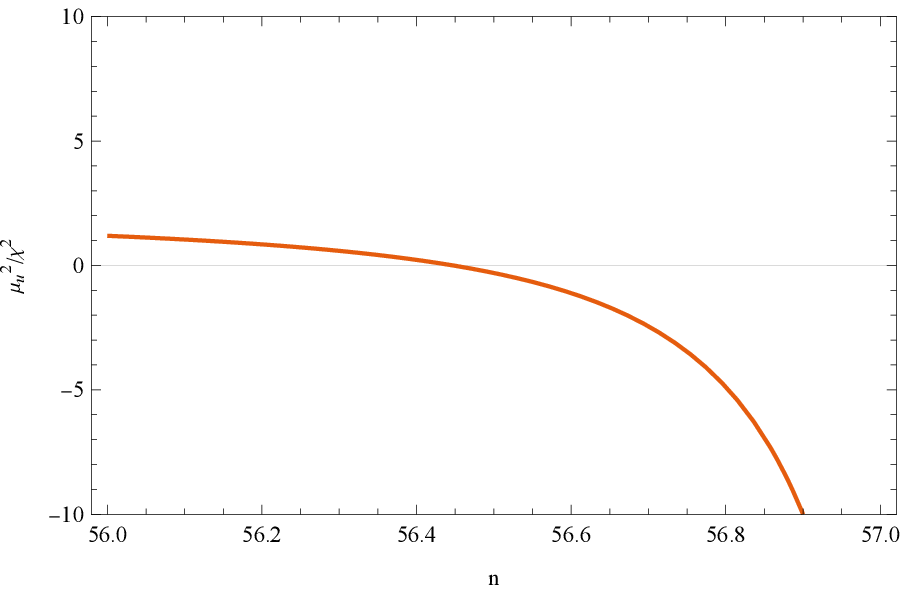}
\includegraphics[width=6.5cm]{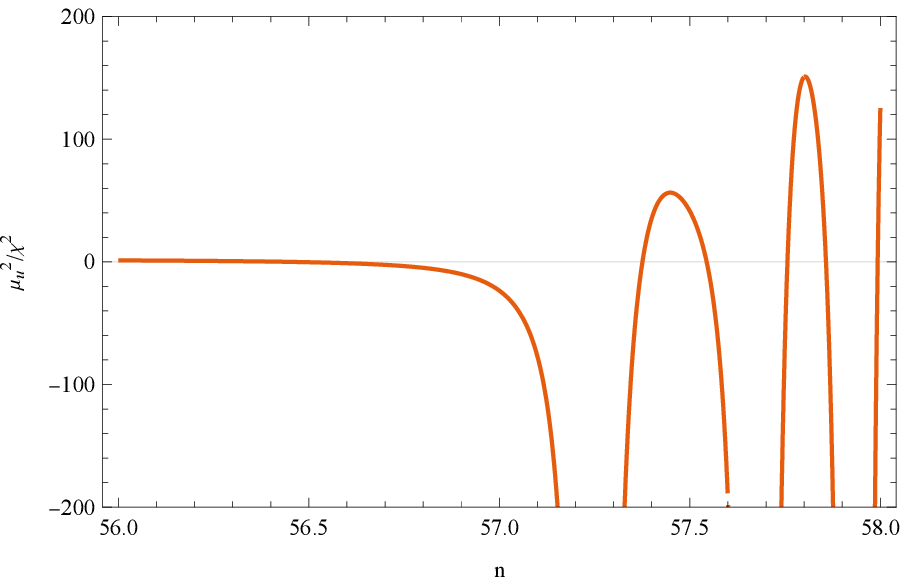}
\caption{\footnotesize The left hand plot shows $\mu_u^2(n)/\chi^2(n)$ 
on the range $56 \leq n \leq 57$, just before the first zero of $\mu(n)$. The 
right hand plot depicts $\mu_u^2(n)/\chi^2(n)$ over the slightly larger range
of $56 \leq n \leq 58$, which includes the first three zeroes of $\mu(n)$.}
\label{Zoomumass}
\end{figure}

The slow roll approximation (\ref{slowroll}) tells us that $\psi' 
\simeq -1/\psi$ and $\chi(n) \simeq c/\sqrt{3} \times \psi(n)$. Setting
$D=4$, and using our values of $c = 7.1 \times 10^{-6}$ and $q = 1.2 
\times 10^{-6}$, gives the mass hierarchy,
\begin{equation}
\frac{\mu^2_u}{\chi^2} \simeq 2 + \frac{6 q^2}{c^2} \simeq 2.16 \quad >
\quad \frac{\mu^2}{\chi^2} \simeq \frac{6 q^2}{c^2} \simeq 0.16 \quad >
\quad \frac{\mu^2_t}{\chi^2} \simeq \frac{3}{\psi^2} \; . \label{masses}
\end{equation}
Figure~\ref{Earlymass} shows the various masses through inflation.
\noindent As one can just see from the larger $n$ values of 
Figure~\ref{Earlymass}, the hierarchy of equation (\ref{masses}) becomes 
inverted after the end of inflation. Figure~\ref{Latemass} shows the 
behavior after the end of inflation. During this phase $\mu^2_u/\chi^2$ 
is mostly tachyonic, and actually diverges at points where $\psi(n) = 0$. 
On the other hand, $\mu^2/\chi^2$ oscillates between $0$ and the small
value of $0.16$, while $\mu^2_t/\chi^2$ grows monotonically to large, 
positive values. The $u$-mode mass is the most important of the three, 
and its evolution is the most complex. Figure~\ref{Zoomumass} shows its 
behavior in more detail.

\subsection{Approximating the Longitudinal Amplitude}

\begin{figure}[H]
\centering
\includegraphics[width=4.3cm]{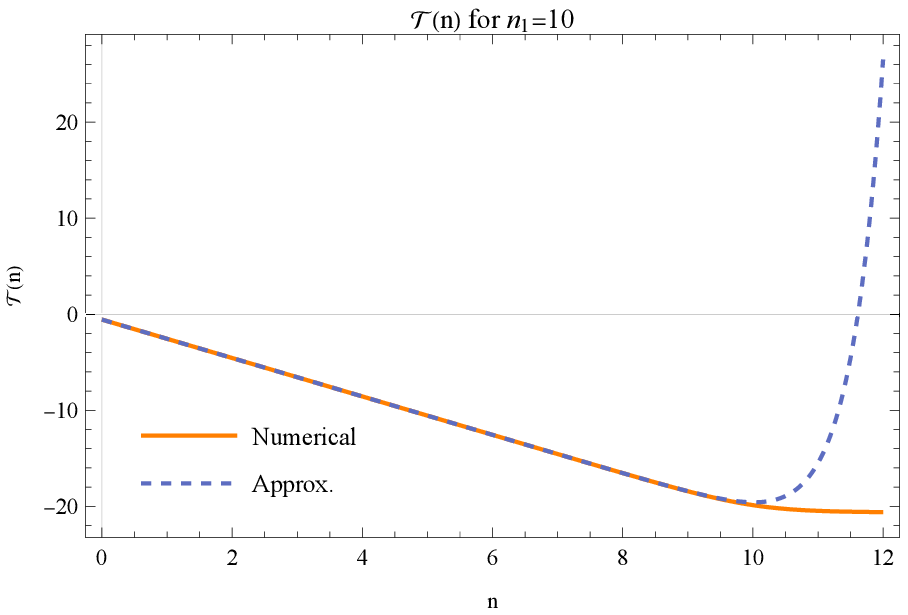}
\includegraphics[width=4.3cm]{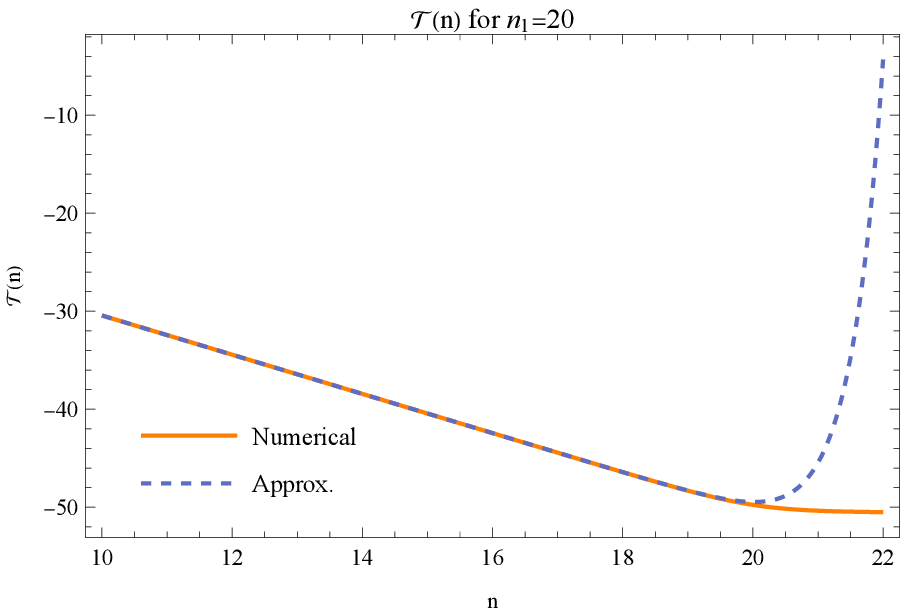}
\includegraphics[width=4.3cm]{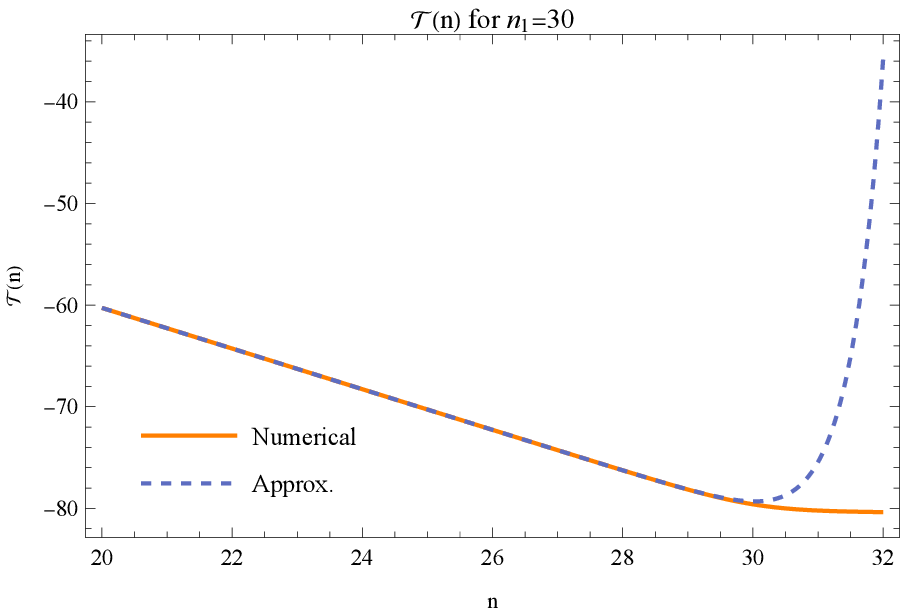}
\caption{\footnotesize Plot comparing $\mathcal{T}(n,\kappa)$ with the
ultraviolet form (\ref{TUVexp}) for modes which experience first horizon
crossing at $n_1 = 10$ (left), $n_1 = 20$ (center) and $n_1 = 30$ (right).}
\label{EarlyT}
\end{figure}

Equation (\ref{Teqn}) for $\mathcal{T}(n,\kappa)$ contains six terms. The 
ultraviolet regime is defined by the condition $\kappa \gg \chi(n) e^{n}$. 
In this regime equation (\ref{Teqn}) is dominated by the 4th and 6th terms, 
$2\kappa^2 e^{-2n}/\chi^2$ and $-e^{-2 [\mathcal{T} + (D-1)n]}/2 \chi^2$, 
and the amplitude takes the form,
\begin{eqnarray}
\lefteqn{\mathcal{T}(n,\kappa) = \ln\Bigl[\frac1{2\kappa}\Bigr] - (D\!-\!2) n }
\nonumber \\
& & \hspace{2.7cm} + \Bigl[ \frac12 (D\!-\!2) (D \!-\! 2\epsilon) - 
\frac{2 \mu_t^2}{\chi^2}\Bigr] \Bigl( \frac{\chi e^n}{2 \kappa}\Bigr)^2 + 
O\Biggl( \Bigl(\frac{\chi e^n}{2 \kappa}\Bigr)^4 \Biggr) \; . \qquad 
\label{TUVexp}
\end{eqnarray}
Figure~\ref{EarlyT} compares numerical evolution of the exact equation 
(\ref{Teqn}) with the ultraviolet form (\ref{TUVexp}) for wave numbers
which experience first horizon crossing at $n_1 = 10$, $n_1 = 20$, and
$n_1 = 30$. The agreement is excellent up to horizon crossing.

\begin{figure}[H]
\centering
\includegraphics[width=4.3cm]{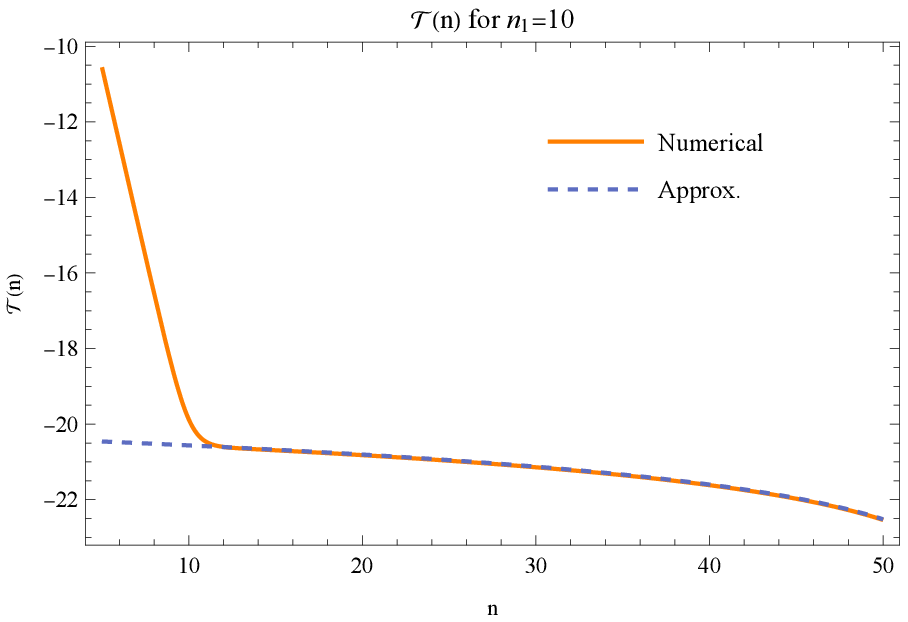}
\includegraphics[width=4.3cm]{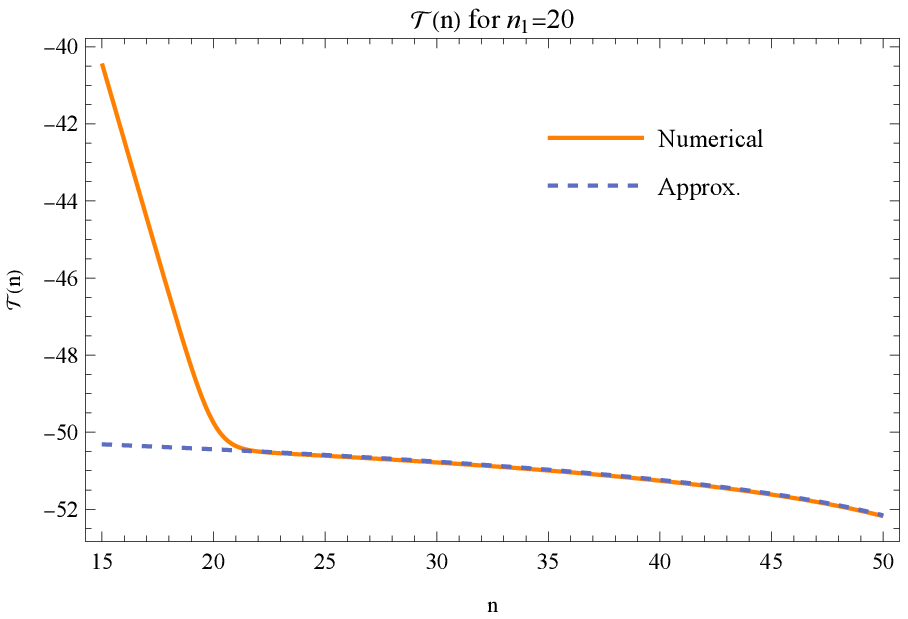}
\includegraphics[width=4.3cm]{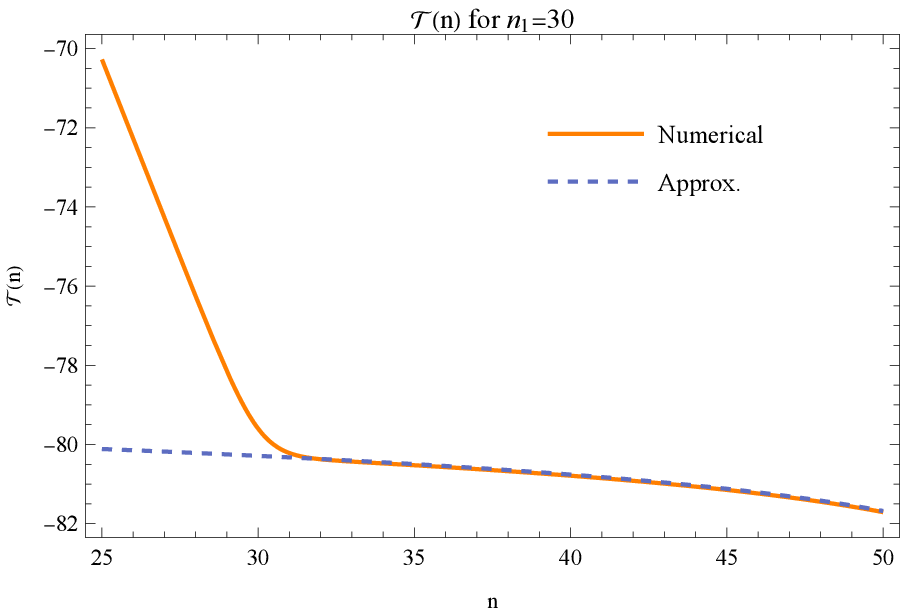}
\caption{\footnotesize Plot comparing $\mathcal{T}(n,\kappa)$ with the
late time form (\ref{TIRform}) for modes which experience first horizon
crossing at $n_1 = 10$ (left), $n_1 = 20$ (center) and $n_1 = 30$ (right).}
\label{LateT}
\end{figure}

After 1st horizon crossing the 4th and 6th terms of (\ref{Teqn}) effectively 
drop out and the relation simplifies to,
\begin{equation}
\mathcal{T}'' + \frac12 {\mathcal{T}'}^2 + (D \!-\! 1 \!-\! \epsilon)
\mathcal{T}' - 2 (D\!-\!1\!-\!\epsilon) \frac{\mu'}{\mu} - 2 \frac{\mu''}{\mu}
\simeq 0 \; . \label{Teqnsimp}
\end{equation}
This is an equation for $\mathcal{T}'$, and it is easy to see that a 
particular solution is,
\begin{equation}
\mathcal{T}' = 2 \frac{\mu'}{\mu} \; . \label{Tprime}
\end{equation}
Integrating (\ref{Tprime}), and using the tensor power spectrum to infer 
the integration constant to all orders in the slow roll approximation 
\cite{Brooker:2015iya}, implies,\footnote{The integration constant in relation
(\ref{TIRform}) suffices for smooth inflationary potentials. When features are
present the constant can be supplemented by known corrections which depend
nonlocally on the expansion history before first horizon crossing 
\cite{Brooker:2017kij}.}
\begin{equation}
\mathcal{T}(n,\kappa) \simeq \ln\Biggl[ \frac{\chi_1^2 C(\epsilon_1)}{2 \kappa^3}
\times \frac{\mu^2(n)}{\mu^2_1} \Biggr] \; , \label{TIRform}
\end{equation}
where the function $C(\epsilon)$ is,
\begin{equation}
C(\epsilon) = \frac1{\pi} \Gamma^2\Bigl( \frac12 \!+\! \frac1{1 \!-\! \epsilon}
\Bigr) \Bigl[2 (1 \!-\! \epsilon)\Bigr]^{\frac2{1-\epsilon}} \; . \label{Cdef}
\end{equation}
Figure~\ref{LateT} compares the exact numerical result with the infrared form
(\ref{TIRform}) for modes which experience horizon crossing at $n_1 = 10$,
$n_1 = 20$, and $n_1 = 30$. Agreement is excellent.

Although one can see from Figure~\ref{LateT} that the approximate solution 
(\ref{TIRform}) is highly accurate, it cannot be exact for two reasons:
\begin{enumerate}
\item{We neglected the 4th and 6th terms in simplifying equation (\ref{Teqn})
to reach (\ref{Teqnsimp}); and}
\item{Just because (\ref{Tprime}) is {\it a} solution to (\ref{Teqnsimp}) does 
not mean it is {\it the} solution.}
\end{enumerate}
To find the {\it general} solution to (\ref{Teqnsimp}) we substitute 
$\mathcal{T}' = 2 \mu'/\mu + f(n)$,
\begin{equation}
f' + 2 \frac{\mu'}{\mu} f + \frac12 f^2 + (D \!-\! 1 \!-\! \epsilon) f \simeq 
0 \; . \label{feqn1}
\end{equation}
Now divide by $\mu^2 e^{(D-1) n} \chi f^2$ to reach the form,
\begin{equation}
\frac{\partial}{\partial n} \Biggl[ \frac1{\mu^2 e^{(D-1) n} \chi f}\Biggr] =
\frac1{2 \mu^2 e^{(D-1)n} \chi} \; . \label{feqn2}
\end{equation}
Integrating equation (\ref{feqn2}) from some point $n_2$ gives the general
solution,
\begin{eqnarray}
\lefteqn{f(n) = f_2 \Biggl[ e^{(D-1)(n-n_2)} \Bigl[ \frac{\chi(n)}{\chi_2}\Bigr] 
\Bigl[ \frac{\mu(n)}{\mu_2}\Bigr]^2 } \nonumber \\
& & \hspace{4cm} + \frac12 f_2 \int_{n_2}^{n} \!\!\!\!\! dn' \, e^{(D-1)(n-n')} 
\Bigl[ \frac{\chi(n)}{\chi(n')}\Bigr] \Bigl[ \frac{\mu(n)}{\mu(n')} \Bigr]^2 
\Biggr]^{-1} \; . \qquad \label{fsol}
\end{eqnarray}

Careful consideration of (\ref{fsol}) reveals that $\mathcal{T}(n,\kappa)$ actually
has a finite limit as the mass vanishes. To see this, assume $n$ is such that 
$\mu(n) \rightarrow 0$, and expand the integral of (\ref{fsol}) for small $\mu(n)$,
\begin{eqnarray}
\lefteqn{\int_{n_2}^{n} \!\!\!\!\! dn' \, e^{(D-1)(n-n')} \Bigl[ 
\frac{\chi(n)}{\chi(n')}\Bigr] \Bigl[ \frac{\mu(n)}{\mu(n')} \Bigr]^2 \!\!= 
-\frac{\mu(n)}{\mu'(n)} } \nonumber \\
& & \hspace{2cm} - \frac12 \Bigl[D\!-\!1 \!-\! \epsilon(n) \!+\! 
\frac{\mu''(n)}{\mu'(n)} \Bigr] \Bigl[ \frac{\mu(n)}{\mu'(n)}\Bigr]^2 \ln[\mu^2(n)] 
+ O(1) \; . \qquad \label{singexp}
\end{eqnarray}
Near the point where $\mu(n) \rightarrow 0$ we therefore have,
\begin{equation}
f(n) \longrightarrow -\frac{2 \mu'(n)}{\mu(n)} + \Bigl[D \!-\! 1 \!-\! \epsilon(n)
\!+\! \frac{\mu''(n)}{\mu'(n)}\Bigr] \ln[\mu^2(n)] + O(1) \; . \label{fexp}
\end{equation}
Hence we have,
\begin{equation}
\mathcal{T}'(n,\kappa) \longrightarrow \Bigl[D\!-\!1 \!-\! \epsilon(n) \!+\!
\frac{\mu''(n)}{\mu'(n)} \Bigr] \ln[\mu^2(n)] + O(1) \; . \label{Tpfull}
\end{equation}
Although expression (\ref{Tpfull}) diverges as $\mu(n)$ goes to zero, the 
singularity is integrable, which means that $\mathcal{T}(n,\kappa)$ remains
finite.
\begin{figure}[H]
\centering
\includegraphics[width=6cm]{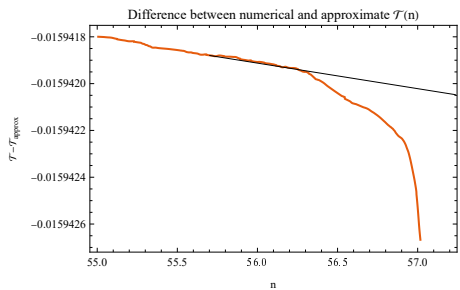}
\caption{\footnotesize Numerical determination of the constant $f_2$ from
the difference of $\mathcal{T}(n,\kappa)$ and expression (\ref{TIRform}).
In this case $\kappa$ was chosen to experience first horizon crossing at
$n_1 = 10$.}
\label{f2plot}
\end{figure} 
As we see from Figure~\ref{f2plot}, the constant $f_2$ in expression 
(\ref{fsol}) represents the difference between the actual value of 
$\mathcal{T}'(n_2,\kappa)$ and its approximate form (\ref{Tprime}) $2 
\mu'(n_2)/\mu(n_2)$. Because the approximate form is quite accurate, $f_2$ 
is a very small number, about $f_2 \sim -10^{-7}$. The fact that $f_2$ drops 
out of the asymptotic form (\ref{fexp}) means that the ultimate finiteness 
of $\mathcal{T}(n,\kappa)$ is a robust conclusion. However, $\mu(n)$ must 
be {\it very} close to zero before the integral (\ref{singexp}) begins to 
dominate over the first term in the denominator of (\ref{fsol}), which has 
a relative enhancement of $e^{(D-1)(n-n_2)}$. If we take $n_2 = 56$ and 
define $n_* \simeq 57.25$ as the first zero of $\mu(n)$, the point $n_f$ at 
which expressions (\ref{fexp}-\ref{Tpfull}) become valid approximately obeys,
\begin{equation}
n_* - n_f \simeq \frac12 f_2 \, e^{-3 (n_* - n_2)} \Bigl( \frac{\chi_2}{\chi_*} 
\Bigr) \Bigl(\frac{\mu_2}{\mu_*'} \Bigr)^2 \simeq 10^{-9} \; . \label{fdomn}
\end{equation}
We can therefore estimate the minimum value of $\mathcal{T}(n,\kappa)$ as
\begin{equation}
\mathcal{T}_{\rm min} \simeq \ln\Biggl[ \frac{\chi_1^2 C(\epsilon_1)}{2 \kappa^3}
\!\times\! \frac14 f_2^2 \, e^{-6 (n_* - n_2)} \Bigl( \frac{\chi_2}{\chi_*}\Bigr)^2
\Bigl( \frac{\mu_2}{\mu_1}\Bigr)^2 \Bigl( \frac{\mu_2}{\mu_*'}\Big)^2 \Biggr] \; . 
\label{Tmin}
\end{equation}

\subsection{Approximating the Temporal Amplitude}

\begin{figure}[H]
\centering
\includegraphics[width=4.3cm]{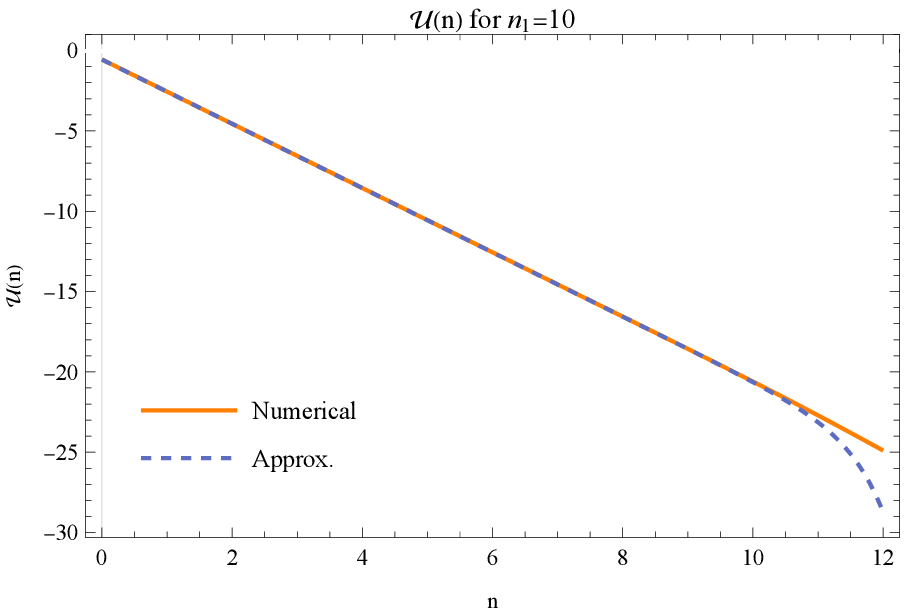}
\includegraphics[width=4.3cm]{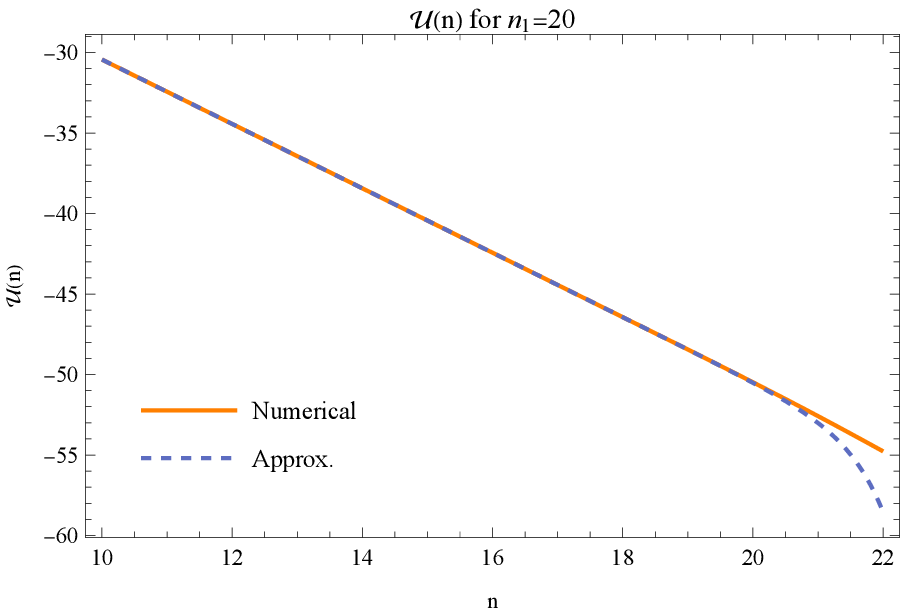}
\includegraphics[width=4.3cm]{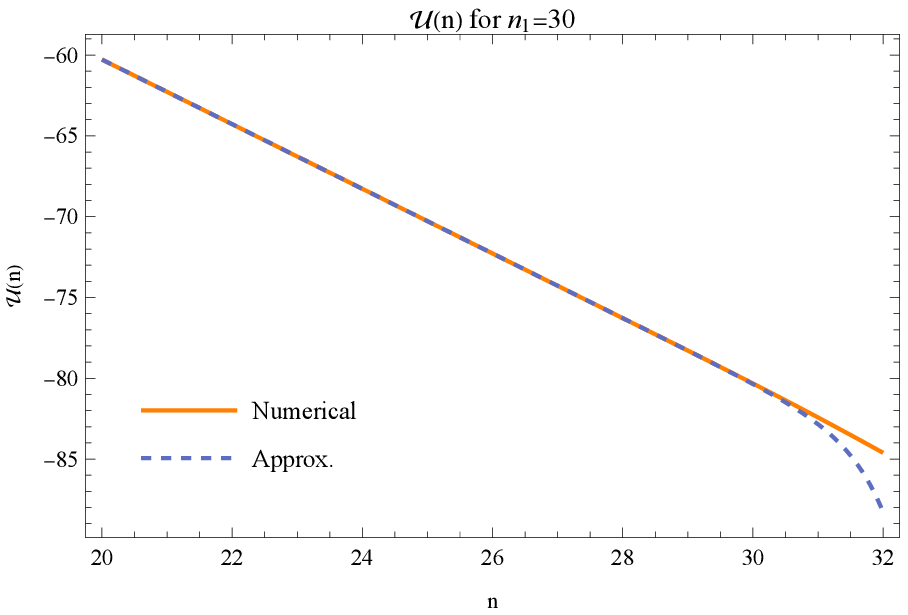}
\caption{\footnotesize Plot comparing $\mathcal{U}(n,\kappa)$ with the
ultraviolet form (\ref{UUVexp}) for modes which experience first horizon
crossing at $n_1 = 10$ (left), $n_1 = 20$ (center) and $n_1 = 30$ (right).}
\label{EarlyU}
\end{figure}

Equation (\ref{Ueqn}) for $\mathcal{U}(n,\kappa)$ contains the same six 
terms as (\ref{Teqn}). In the ultraviolet it is also dominated by the 4th 
and 6th terms, $2\kappa^2 e^{-2n}/\chi^2$ and $-e^{-2 [\mathcal{U} + 
(D-1)n]}/2 \chi^2$. Hence the ultraviolet expansion of $\mathcal{U}(n,\kappa)$
takes the same form as (\ref{TUVexp}),
\begin{eqnarray}
\lefteqn{\mathcal{U}(n,\kappa) = \ln\Bigl[\frac1{2\kappa}\Bigr] - (D\!-\!2) n }
\nonumber \\
& & \hspace{2.7cm} + \Bigl[ \frac12 (D\!-\!2) (D \!-\! 2\epsilon) - 
\frac{2 \mu_u^2}{\chi^2}\Bigr] \Bigl( \frac{\chi e^n}{2 \kappa}\Bigr)^2 + 
O\Biggl( \Bigl(\frac{\chi e^n}{2 \kappa}\Bigr)^4 \Biggr) \; . \qquad 
\label{UUVexp}
\end{eqnarray}
Figure~\ref{EarlyU} compares the exact numerical solution with the ultraviolet
form (\ref{UUVexp}) for modes which experience first horizon crossing at $n_1 
= 10$, $n_1 = 20$ and $n_1 = 30$. Agreement is excellent up to first horizon 
crossing, just as it was in the analogous comparison of Figure~\ref{EarlyT} for 
$\mathcal{T}(n,\kappa)$.

\begin{figure}[H]
\centering
\includegraphics[width=4.3cm]{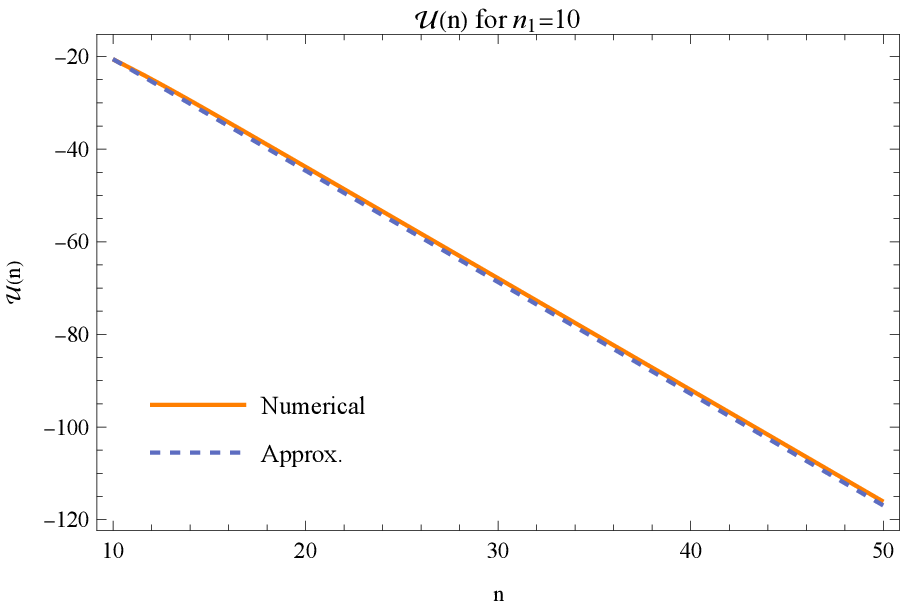}
\includegraphics[width=4.3cm]{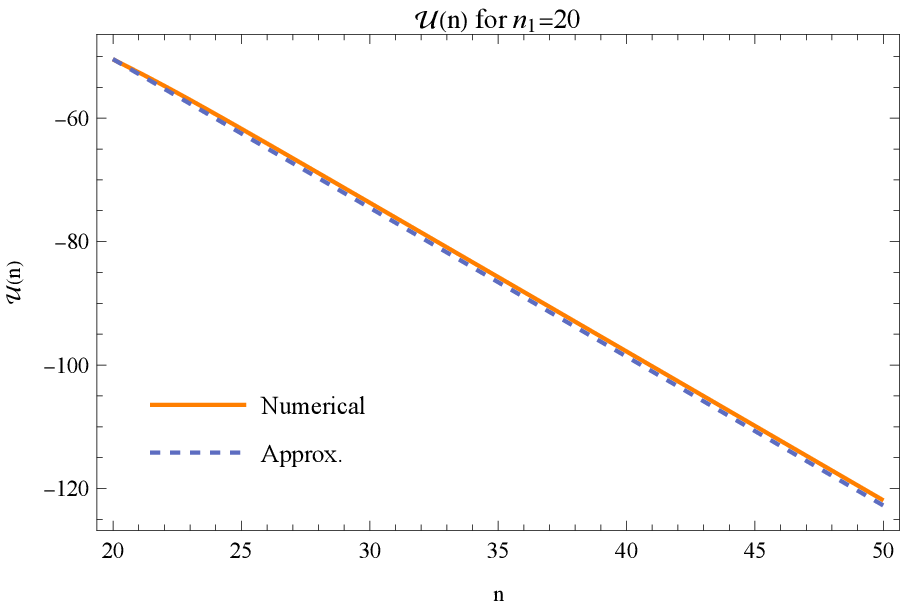}
\includegraphics[width=4.3cm]{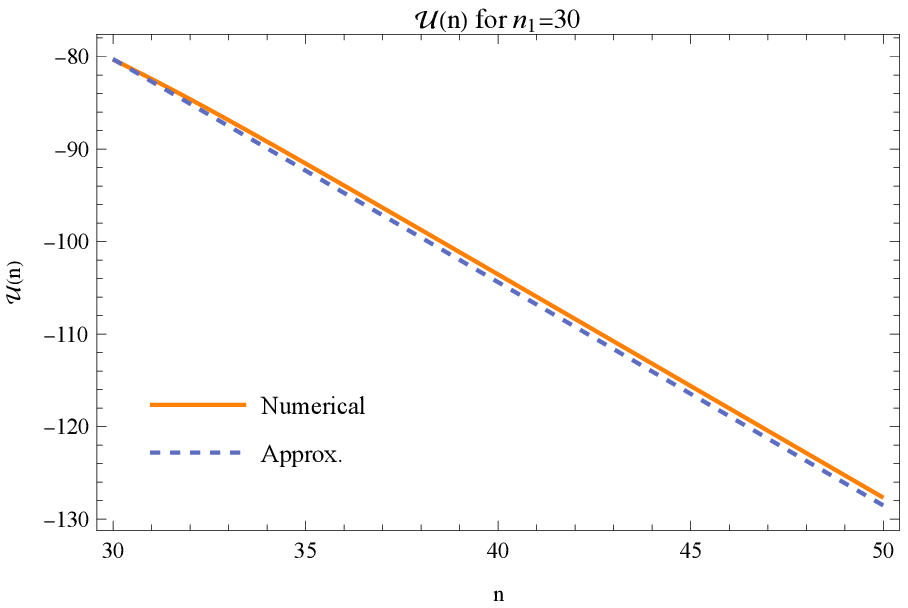}
\caption{\footnotesize Plot comparing $\mathcal{U}(n,\kappa)$ with the
late time form (\ref{UIRform}) for modes which experience first horizon
crossing at $n_1 = 10$ (left), $n_1 = 20$ (center) and $n_1 = 30$ (right).}
\label{LateU}
\end{figure}

The 4th and 6th terms of (\ref{Ueqn}) drop out after first horizon crossing,
and the relation simplifies to,
\begin{equation}
\mathcal{U}'' + \frac12 {\mathcal{U}'}^2 + (D\!-\!1\!-\! \epsilon) \mathcal{U}'
+ \frac{2 \mu^2_u}{\chi^2} \simeq 0 \; . \label{Ueqnsimp}
\end{equation}
Recall from Figure~\ref{Earlymass} that $\mu_u^2(n)/\chi^2(n)$ is approximately
constant during inflation. This means that equation (\ref{Ueqnsimp}) can be 
roughly solved as,
\begin{equation}
\mathcal{U}'(n,\kappa) \simeq -(D\!-\!1\!-\!\epsilon) + \sqrt{(D \!-\! 1 \!-\!
\epsilon)^2 - \frac{4 \mu_u^2}{\chi^2}} \; . \label{approxUprime}
\end{equation}
With the appropriate integration constant we therefore have,
\begin{eqnarray}
\lefteqn{\mathcal{U}(n,\kappa) \simeq \ln\Biggl[ \frac{\chi_1^2 C(\epsilon_1)}{
2 \kappa^3} \!\times\! \frac{\chi_1}{\chi(n)} \Biggl] - (D \!-\! 1) (n \!-\! n_1)
} \nonumber \\
& & \hspace{5cm} + \int_{n_1}^{n} \!\!\!\! dn' \sqrt{[D\!-\! 1 \!-\! \epsilon(n')]^2
- \frac{4 \mu_u^2(n')}{\chi^2(n')} } \; . \qquad \label{UIRform}
\end{eqnarray}
Figure~\ref{LateU} compares this approximation with the numerical evolution
for modes which experience horizon crossing at $n_1 = 10$, $n_1 = 20$ and
$n_1 = 30$. 

After the end of inflation $\mu^2(n)/\chi^2(n)$ falls off whereas the 
derivative terms in $\mu^2_u/\chi^2$ become large and tachyonic. This means
we can neglect $\mu^2(n)/\chi^2(n)$,
\begin{equation}
\frac{\mu_u^2}{\chi^2} \simeq (D\!-\!2) (1\!-\! \epsilon) - (D\!-\!3\!+\!
\epsilon) \frac{\mu'}{\mu} + \Bigl( \frac{\mu'}{\mu}\Bigr)' - \Bigl(
\frac{\mu'}{\mu}\Bigr)^2 \; . \label{umasssimp2}
\end{equation}
We now make the substitution,
\begin{equation}
\mathcal{U}'(n,\kappa) = -\frac{2 \mu'(n)}{\mu(n)} - 2 (D\!-\!2) + g(n) \; ,
\label{Uansatz}
\end{equation}
in equation (\ref{Ueqnsimp}) to find,
\begin{equation}
g' - \Bigl(D \!-\! 3 \!+\! \epsilon \!+\! 2 \frac{\mu'}{\mu}\Bigr) g 
+ \frac12 g^2 = 0 \; . \label{geqn}
\end{equation}
Multiplying by $e^{(D-3)n} \mu^2(n)/[\chi(n) g^2(n)]$ makes the $g$-dependent
terms a total derivative, and permits us to write the general solution as,
\begin{eqnarray}
\lefteqn{ g(n) = g_2 \Biggl[ e^{-(D-3) (n-n_2)} \Bigl[ \frac{\chi(n)}{\chi_2}
\Bigr] \Bigl[ \frac{\mu_2}{\mu(n)}\Bigr]^2 } \nonumber \\
& & \hspace{4cm} + \frac12 g_2 \!\! \int_{n_2}^{n} \!\!\!\!\! dn' \, 
e^{-(D-3) (n-n')} \Bigl[ \frac{\chi(n)}{\chi(n')}\Bigr] \Bigl[ \frac{\mu(n')}{
\mu(n)}\Bigr]^2 \Biggr]^{-1} \; , \qquad \label{gsol}
\end{eqnarray} 
where the constant $g_2$ is determined to interpolate between (\ref{approxUprime})
and (\ref{Uansatz}),
\begin{equation}
g_2 = \frac{2 \mu_2'}{\mu_2} + (D\!-\!3\!+\!\epsilon_2) +
\sqrt{(D\!-\! 1 \!-\! \epsilon_2)^2 - \frac{4 \mu_2^2}{\chi_2^2}} \; . 
\label{g2def}
\end{equation}
Note that, whereas $f(n)$ diverges as $\mu(n)$ approaches zero, $g(n)$ goes to
zero like $\mu^2(n)$.

\begin{figure}[H]
\centering
\includegraphics[width=4.3cm]{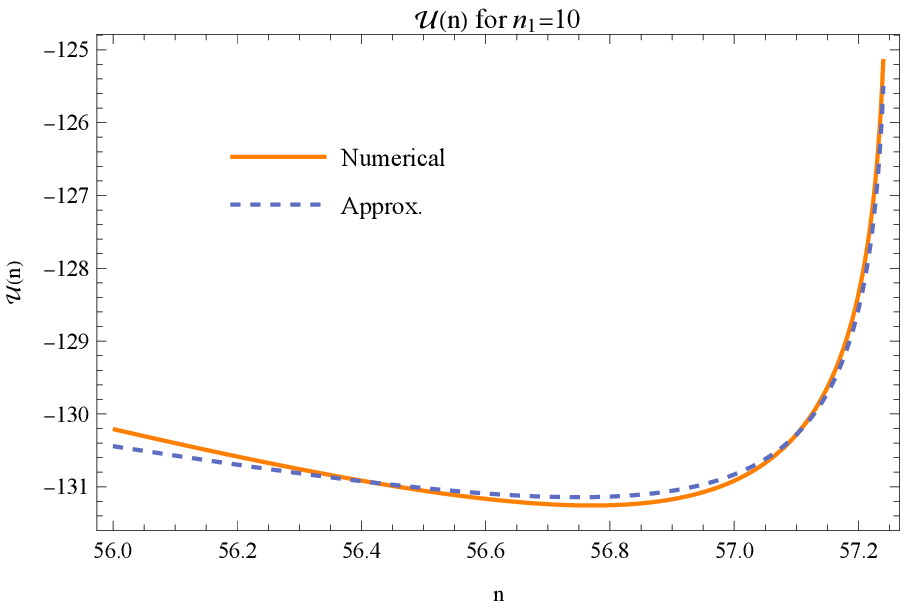}
\includegraphics[width=4.3cm]{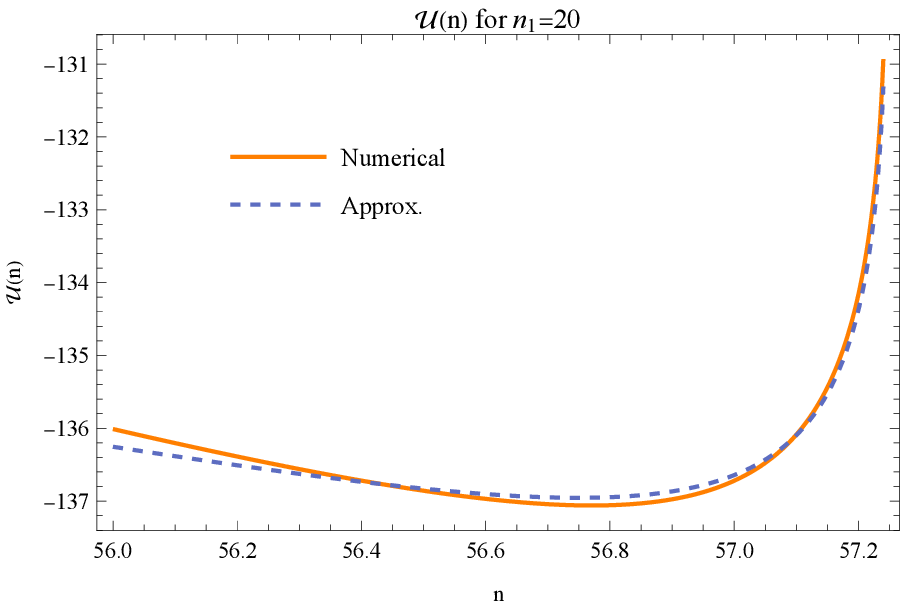}
\includegraphics[width=4.3cm]{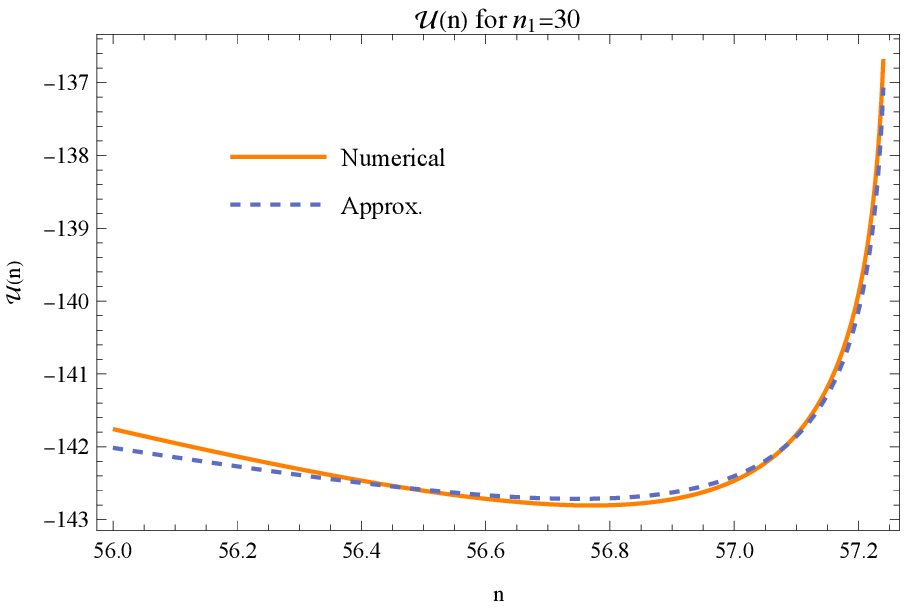}
\caption{\footnotesize Plot comparing $\mathcal{U}(n,\kappa)$ with the
post-inflationary form (\ref{UPostform}) for modes which experience first horizon
crossing at $n_1 = 10$ (left), $n_1 = 20$ (center) and $n_1 = 30$ (right).}
\label{UPost}
\end{figure}

Integrating equation (\ref{Uansatz}), and using (\ref{UIRform}) to supply the
integration constant, gives,
\begin{eqnarray}
\lefteqn{ \mathcal{U}(n,\kappa) \simeq \ln\Biggl[ \frac{\chi_1^2 C(\epsilon_1)}{
2 \kappa^3} \!\times\! \frac{\chi_1}{\chi_2} \!\times\! \frac{\mu_2^2}{\mu^2(n)}
\Biggr] - (D\!-\!1) (n_2 \!-\! n_1) - 2 (D\!-\!2) (n \!-\! n_2) } \nonumber \\
& & \hspace{3.2cm} + \int_{n_1}^{n_2} \!\!\!\!\! dn' \sqrt{ [D\!-\! 1 \!-\! 
\epsilon(n')]^2 - \frac{4 \mu_u^2(n')}{\chi^2(n')} } + \int_{n_2}^{n} \!\!\!\!\!
dn' \, g(n') \; . \qquad \label{UPostform}
\end{eqnarray}
Because $g(n)$ vanishes as $\mu(n) \rightarrow 0$, the $-\ln[\mu^2(n)]$ 
divergence of $\mathcal{U}(n,\kappa)$ is robust. Note that this is not even 
affected by neglecting $\mu^2(n)/\chi^2(n)$ in (\ref{umasssimp2}). 
Figure~\ref{UPost} compares the numerical solution with our analytic 
approximation (\ref{UPostform}).

\section{Quantum-Correcting the Inflaton $0$-Mode}

The purpose of this section is to use the photon propagator to quantum-correct 
the classical equation for the inflaton $0$-mode from (\ref{0order}) to,
\begin{equation}
\partial_0 \Bigl[a^{D-2} \partial_0 \varphi_0\Bigr] + a^{D} \varphi_0
V'(\varphi_0^2) + q^2 \varphi_0 a^{D-2} \eta^{\mu\nu} i \Bigl[\mbox{}_{\mu}
\Delta_{\nu}\Bigr](x;x) = 0 \; . \label{new0modeeqn}
\end{equation}
We begin by deriving exact expressions for the $t$-mode and $u$-mode contributions
to the trace of the photon propagator. We then use a variant of the work-energy theorem
to show how reheating occurs. 

\subsection{The Effective Force}

The $t$-mode contribution to the coincidence limit of the trace of the photon
propagator in equation (\ref{new0modeeqn}) is,
\begin{equation}
\sqrt{-g} \, g^{\mu\nu} i\Bigl[ \mbox{}_{\mu} \Delta_{\nu}\Bigr]_{t}(x;x) =
a^{D-2} \!\! \int \!\! \frac{d^{D-1} k}{(2\pi)^{D-1}} \Biggl\{ \frac1{M^2} \,
\widehat{\partial}_0 t \!\cdot\! \widehat{\partial_0} t^* - \frac{k^2}{M^2}
\, t \!\cdot \! t^* \Biggr\} \; . \label{ttrace1}
\end{equation}
The $t$-mode equation (\ref{teqn}) can be exploited to write the product of 
time derivatives in terms of the norm-squared,
\begin{equation}
\partial_0 t \!\cdot\! \partial_0 t^* = \Bigl[ \frac12 \partial_0^2 + \frac12
(D\!-\!2) a H \partial_0 + k^2 - (D\!-\!2) a H \frac{\partial_0 M}{M} -
\frac{\partial_0^2 M}{M} \Bigr] (t t^*) \; . \label{tID1}
\end{equation}
Using this identity we can re-express the $t$-mode contribution (\ref{ttrace1})
as,
\begin{eqnarray}
\lefteqn{\sqrt{-g} \, g^{\mu\nu} i\Bigl[ \mbox{}_{\mu} \Delta_{\nu}\Bigr]_{t} = 
\frac{a^{D-2}}{M^2} \!\! \int \!\! \frac{d^{D-1}k}{(2\pi)^{D-1}} \Biggl\{ 
\frac12 \partial_0^2 + \frac12 (D\!-\!2) a H \partial_0 - \frac{\partial_0 M}{M}
\partial_0 } \nonumber \\
& & \hspace{4cm} - (D\!-\!2) a H \frac{\partial_0 M}{M} - \frac{\partial_0^2 M}{M}
+ \Bigl( \frac{\partial_0 M}{M}\Bigr)^2 \Biggr\} (t t^*) \; . \qquad 
\label{ttrace2}
\end{eqnarray}
Converting to dimensionless form, and employing equation (\ref{Teqn}) to eliminate
second derivatives, gives the compact form,
\begin{eqnarray}
\lefteqn{\sqrt{-g} \, g^{\mu\nu} i\Bigl[ \mbox{}_{\mu} \Delta_{\nu}\Bigr]_{t}(x;x) 
= \frac{e^{D n} \chi^2(n)}{(8\pi G)^{\frac{D}2 -1} \mu^2(n)} \!\! \int \!\! 
\frac{d^{D-1} \kappa}{(2\pi)^{D-1}} \, e^{\mathcal{T}(n,\kappa)} } \nonumber \\
& & \hspace{1.3cm} \times \Biggl\{ \Bigl[ \frac12 \mathcal{T}'(n,\kappa) - 
\frac{\mu'(n)}{\mu(n)} \Bigr]^2 - \frac{\kappa^2 e^{-2n}}{\chi^2(n)} + 
\frac1{4 \chi^2} e^{-2 [\mathcal{T}(n,\kappa) + (D-1)n]} \Biggr\} . \qquad 
\label{ttrace3}
\end{eqnarray}

The $u$-mode contribution to the photon trace in (\ref{new0modeeqn}) is,
\begin{equation}
\sqrt{-g} \, g^{\mu\nu} i\Bigl[ \mbox{}_{\mu} \Delta_{\nu}\Bigr]_{u}(x;x) =
a^{D-2} \!\! \int \!\! \frac{d^{D-1} k}{(2\pi)^{D-1}} \Biggl\{ -\frac{k^2}{M^2} 
\, u \!\cdot\! u^* + \frac{1}{M^2} \widetilde{D} u \!\cdot\! \widetilde{D} u^*
\Biggr\} \; . \label{utrace1}
\end{equation}
We can eliminate the norm-square of $\partial_0 u$ using the $u$-mode equation
(\ref{ueqn}),
\begin{eqnarray}
\lefteqn{\partial_0 u \!\cdot\! \partial_0 u^* = \Biggl[ \frac12 \partial_0^2 + 
\frac12 (D\!-\!2) a H \partial_0 + k^2 + a^2 M^2 + (D\!-\!2) a^2 H^2 (1 \!-\! 
\epsilon) } \nonumber \\
& & \hspace{3.2cm} - (D\!-\!2) a H \frac{\partial_0 M}{M} + \partial_0 \Bigl( 
\frac{\partial_0 M}{M}\Bigr) - \Bigl(\frac{\partial_0 M}{M} \Bigr)^2 \Biggr] 
(u u^*) \; . \qquad \label{uID1}
\end{eqnarray}
Substituting (\ref{uID1}) in (\ref{utrace1}), and taking apart the factors of
$\widetilde{D} = \partial_0 + (D-2) a H + \partial_0 M/M$ gives,
\begin{eqnarray}
\lefteqn{\sqrt{-g} \, g^{\mu\nu} i\Bigl[ \mbox{}_{\mu} \Delta_{\nu}\Bigr]_{u} = 
\frac{a^{D-2}}{M^2} \!\! \int \!\! \frac{d^{D-1}k}{(2\pi)^{D-1}} \Biggl\{ 
\frac12 \partial_0^2 + \frac32 (D\!-\!2) a H \partial_0 + \frac{\partial_0 M}{M}
\partial_0 + a^2 M^2 } \nonumber \\
& & \hspace{.7cm} + (D\!-\!2) (D\!-\!1\!-\!\epsilon) a^2 H^2 + (D\!-\!2) a H 
\frac{\partial_0 M}{M} + \partial_0 \Bigl(\frac{\partial_0 M}{M} \Bigr) 
\Biggr\} (u u^*) \; . \qquad \label{utrace2}
\end{eqnarray}
The final, dimensionless form is very similar to (\ref{ttrace3}),
\begin{eqnarray}
\lefteqn{\sqrt{-g} \, g^{\mu\nu} i\Bigl[ \mbox{}_{\mu} \Delta_{\nu}\Bigr]_{u}(x;x) 
= \frac{e^{D n} \chi^2(n)}{(8\pi G)^{\frac{D}2 -1} \mu^2(n)} \!\! \int \!\! 
\frac{d^{D-1} \kappa}{(2\pi)^{D-1}} \, e^{\mathcal{U}(n,\kappa)} } \nonumber \\
& & \hspace{-0.1cm} \times \Biggl\{ \Bigl[ \frac12 \mathcal{U}'(n,\kappa) + 
\frac{\mu'(n)}{\mu(n)} + D\!-\!2 \Bigr]^2 - \frac{\kappa^2 e^{-2n}}{\chi^2(n)} 
+ \frac1{4 \chi^2} e^{-2 [\mathcal{U}(n,\kappa) + (D-1)n]} \Biggr\} . 
\qquad \label{utrace3}
\end{eqnarray}

\subsection{Reheating}

The dimensionless form of the inflaton $0$-mode equation (\ref{new0modeeqn}) 
takes the form,
\begin{equation}
e^{n} \chi \frac{\partial}{\partial n} \Bigl[ e^{(D-1) n} \chi \psi'\Bigr] =
- e^{D n} \psi \Biggl[ U'(\psi^2) + \frac{Q^2 \chi^2}{\mu^2} \!\! \int \!\!
\frac{d^{D-1}\kappa}{(2\pi)^{D-1}} \Biggl\{ \qquad \Biggr\} \Biggr] \equiv
\mathcal{F} \; . \label{force}
\end{equation}
where the term inside the curly brackets is the sum of the $t$ and $u$ contributions
from expressions (\ref{ttrace3}) and (\ref{utrace3}), and $Q^2 \equiv q^2/(8\pi G)^{
\frac{D}2 -2}$ is the dimensionless charge. Multiplying both sides of (\ref{force})
by $e^{(D-2)n} \chi \psi'$ and integrating gives a curious generalization of the
famous work-energy theorem of introductory physics,
\begin{equation}
e^{(D-1)n} \chi \psi' \frac{\partial}{\partial n} \Bigl[ e^{(D-1) n} \chi \psi'
\Bigr] = \frac12 \frac{\partial}{\partial n} \Bigl[ e^{(D-1) n} \chi \psi'\Bigr]^2
= e^{(D-2) n} \chi \psi' \!\times\! \mathcal{F} \; . \label{WEthm1}
\end{equation}
We now integrate (\ref{WEthm1}) from the beginning of reheating (at $n = n_i$) to
the end (at $n = n_f$), and use the fact that $\psi'(n_f) = 0$ at the end of 
reheating,
\begin{equation}
0 - \frac12 \Bigl[ e^{(D-1) n_i} \chi_i \psi'_i\Bigr]^2 = \int_{n_i}^{n_f} \!\!\!\!
dn \, e^{(D-2) n} \chi(n) \psi'(n) \mathcal{F}(n) \; . \label{WEthm2}
\end{equation}

Equation (\ref{WEthm2}) clearly implies that the product of $\psi'(n) \times
\mathcal{F}(n)$ must be {\it negative} in order to suck the energy out of the 
inflaton $0$-mode. The classical contribution from $\psi' \times -\psi U'(\psi^2)$
is positive, so reheating must be driven by the quantum corrections from the
$t$-modes and the $u$-modes, each of which has two positive and one negative
contribution. From expressions (\ref{ttrace3}) and (\ref{utrace3}) we see that
the desired negative contribution can only come from the $-\kappa^2 
e^{-2n}/\chi^2(n)$ terms, however, it is not clear if the dominant effect
comes from $t$-modes or $u$-modes. It is also not clear whether the largest
contributions come from super-horizon modes (with $\kappa < \chi(n_e) e^{n_e}$,
where $n_e$ denotes the end of inflation) or sub-horizon modes (with $\kappa >
\chi(n_e) e^{n_e}$). Note that discretization can only recover the longest
wavelength sub-horizon modes.

Let us first examine sub-horizon modes, for which $\chi e^{n}/\kappa$ is 
small. In this case the ultraviolet expansions (\ref{TUVexp}) and (\ref{UUVexp})
imply that the multiplicative exponentials agree to leading order,
\begin{equation}
e^{\mathcal{T}(n,\kappa)} = \frac{e^{-(D-2) n}}{2 \kappa} \Biggl\{ 1 + 
O\Bigl( \frac{\chi^2 e^{2n}}{\kappa^2}\Bigr) \Biggr\} \;\; , \;\;
e^{\mathcal{U}(n,\kappa)} = \frac{e^{-(D-2) n}}{2 \kappa} \Biggl\{ 1 + 
O\Bigl( \frac{\chi^2 e^{2n}}{\kappa^2}\Bigr) \Biggr\} \; .
\end{equation}
Substituting the same ultraviolet expansions into the curly bracketed parts
of (\ref{ttrace3}) and (\ref{utrace3}) gives,
\begin{eqnarray}
\lefteqn{ \Bigl[ \frac12 \mathcal{T}' - \frac{\mu'}{\mu}\Bigr]^2 - 
\frac{\kappa^2 e^{-2n}}{\chi^2} + \frac{e^{-2 [\mathcal{T} + (D-1) n]}}{4 \chi^2} 
} \nonumber \\
& & \hspace{2cm} = -\frac12 (D\!-\!2) (1 \!-\! \epsilon) - (1\!-\!\epsilon) 
\frac{\mu'}{\mu} - \Bigl( \frac{\mu'}{\mu}\Bigr)' + O \Bigl( \frac{\chi^2 e^{2n}}{
\kappa^2}\Bigr) \; , \qquad \\
\lefteqn{ \Bigl[ \frac12 \mathcal{U}' + \frac{\mu'}{\mu}+ D\!-\!2 \Bigr]^2 - 
\frac{\kappa^2 e^{-2n}}{\chi^2} + \frac{e^{-2 [\mathcal{U} + (D-1) n]}}{4 \chi^2} 
} \nonumber \\
& & \hspace{2cm} = \frac12 (D\!-\!2) (1 \!-\! \epsilon) + (1\!-\!\epsilon) 
\frac{\mu'}{\mu} + \Bigl( \frac{\mu'}{\mu}\Bigr)' + O \Bigl( \frac{\chi^2 e^{2n}}{
\kappa^2}\Bigr) \; . \qquad
\end{eqnarray}
Hence there is perfect cancellation between the sub-horizon $t$-mode and $u$-mode
contributions at leading order.

Super-horizon modes cannot show the same cancellation because 
$\mathcal{T}(n,\kappa)$ approaches a large, negative constant (\ref{Tmin}) as 
$\mu(n)$ goes to zero, whereas $\mathcal{U}(n,\kappa)$ diverges like 
$\mathcal{U}_* + \ln[\mu_2^2/\mu^2(n)]$.\footnote{The constant $\mathcal{U}_*$ 
can be found from expression (\ref{UPostform}) by extracting the factor of
$\ln[\mu_2^2/\mu^2(n)]$ and then setting $n = n_*$ in the remainder.} This means 
that the multiplicative exponentials take the form,
\begin{equation}
e^{\mathcal{T}(n,\kappa)} \longrightarrow e^{\mathcal{T}_{\rm min}} \qquad , \qquad
e^{\mathcal{U}(n,\kappa)} \longrightarrow e^{\mathcal{U}_*} \Bigl[ \frac{\mu_2}{\mu(n)}
\Bigr]^2 \; . 
\end{equation}
The curly bracketed terms which involve explicit factors of $\mu'/\mu$ wind up
depending on the functions $f(n)$ and $g(n)$, given in expressions (\ref{fsol}) 
and (\ref{gsol}), respectively,
\begin{eqnarray}
\Bigl[ \frac12 \mathcal{T}' - \frac{\mu'}{\mu}\Bigr]^2 &\!\!\! = \!\!\!& \frac14
f^2(n) \longrightarrow \Bigl[\frac{\mu'(n)}{\mu(n)}\Bigr]^2 \; , \\
\Bigl[ \frac12 \mathcal{U}' + \frac{\mu'}{\mu}+ D\!-\!2 \Bigr]^2 &\!\!\! = \!\!\!&
\frac14 g^2(n) \longrightarrow 0 \; .
\end{eqnarray}
This means that the $t$-modes contribute positively, while the $u$-modes make
a negative contribution,
\begin{eqnarray}
e^{\mathcal{T}} \!\times\! \Biggl\{ \qquad \Biggr\} &\!\!\! \longrightarrow \!\!\!&
e^{\mathcal{T}_{\rm min}} \Bigl[ \frac{\mu'(n)}{\mu(n)}\Bigr]^2 \; , \label{tlead} \\
e^{\mathcal{U}} \!\times\! \Biggl\{ \qquad \Biggr\} &\!\!\! \longrightarrow \!\!\!&
e^{\mathcal{U}_*} \Bigl[ \frac{\mu_2}{\mu(n)} \Bigr]^2 \!\times\! -\frac{\kappa^2
e^{-2n_*}}{\chi_*^2} \; . \qquad \label{ulead}
\end{eqnarray}
How large the relative coefficients are depends on the integration constant $f_2$,
for which we do not yet have an analytic form.

Whether (\ref{tlead}) or (\ref{ulead}) dominates, it is significant that both terms
diverge like $1/\mu^2(n)$. Because the effective force contains another factor of 
$1/\mu^2(n)$, this means that the quantum correction diverges like $1/\mu^4(n)$ near
the point $n_*$ at which $\mu(n)$ vanishes. The measure factor in (\ref{WEthm2})
softens this somewhat, but not enough,
\begin{equation}
Q^2 \psi'(n) \psi(n) dn = \frac14 d\mu^2 \; . \label{measure}
\end{equation} 
The integral (\ref{WEthm2}) therefore diverges before $\mu(n) = 0$, which
presumably brings reheating to an end. 

\section{Conclusions}

Ema et al. have shown that coupling a charged inflaton to electromagnetism 
provides the most efficient reheating \cite{Ema:2016dny}. The mechanism is
that the inflaton's evolution induces a time-dependent photon mass through
the Higgs mechanism. Nothing special changes about the transverse spatial
polarizations, but inverse powers of the mass appear in the 
longitudinal-temporal polarizations (\ref{tAs}) and (\ref{uAs}), which result 
from the photon having ``eaten'' the phase of the inflaton field. These 
factors diverge when the inflaton passes through zero. The effect is 
strengthened by factors of $\mu'(n)/\mu(n)$ which appear in the mass terms 
(\ref{tmass}-\ref{umass}) of the two modes.

Our paper represents an effort to improve on previous excellent numerical 
studies of this process based on discretizing space \cite{Bezrukov:2020txg}. 
Although that method can accommodate arbitrarily strong photon fields, it is of 
course limited to a finite range of sub-horizon modes. In contrast, we use the 
trace of the coincident photon propagator to study the inflaton $0$-mode 
equation (\ref{new0modeeqn}). Our expressions (\ref{ttrace2}-\ref{ttrace3}) and 
(\ref{utrace2}-\ref{utrace3}) for the longitudinal and temporal contributions 
to this trace are exact. They can be used to include the effects of super-horizon
modes, and of arbitrarily short wave length modes. In fact, our use of dimensional
regularization means that the far ultraviolet can be included as well, through the
use of expansions (\ref{TUVexp}) and (\ref{UUVexp}).
 
We have also derived good analytic approximations for the amplitudes. Before
first horizon crossing these are (\ref{TUVexp}) and (\ref{UUVexp}), respectively.
After first crossing the $t$-mode amplitude is well approximated by expression
(\ref{TIRform}) until close to the point at which $\mu(n)=0$. However, expression 
(\ref{Tpfull}) shows that the $t$-mode amplitude remains finite when $\mu(n) = 0$. 

Two forms are required to approximate the $u$-mode amplitude after first
horizon crossing, owing to its dependence on the complicated behavior of the 
$u$-mode mass term (\ref{umass}), which is evident from Figures~\ref{Latemass} 
and \ref{Zoomumass}. During inflation, the near constancy of $\epsilon(n)$
and $\mu^2_{u}(n)/\chi^2(n)$, result in expression (\ref{UIRform}) giving a
good approximation. After the end of inflation the better approximation is
provided by expression (\ref{UPostform}). Because this last form becomes exact
as $\mu(n) \rightarrow 0$, we know that the $u$-mode amplitude diverges like
$-\ln[\mu^2(n)]$, which provides an extra factor of $1/\mu^2(n)$ in the trace
of the photon propagator (\ref{utrace3}).

The obvious next step is to exploit the powerful analytic expressions we have 
derived to make a detailed numerical study of reheating in a realistic model,
such as Starobinsky inflation \cite{Starobinsky:1980te}, Higgs inflation 
\cite{Bezrukov:2007ep}, or a hybrid model \cite{Bezrukov:2020txg}. Such an
analysis would begin by renormalizing equation (\ref{new0modeeqn}), and then
focus on determining whether the dominant effect for $\mu(n) \rightarrow 0$ 
comes from sub-horizon or super-horizon modes, and whether it is the $t$-modes
or the $u$-modes which contribute more strongly. Another key issue is whether
or not the effect is so strong that the inflaton is precluded from making 
even a single oscillation. Right now, it seems as if the strongest effect 
comes from super-horizon $u$-modes, and this contribution is so strong that 
the inflaton $0$-mode is prevented from passing through zero.

Finally, our extension of the vector propagator to include time-dependent 
masses in cosmological backgrounds has two obvious applications in addition 
to reheating. The first of these is the study of quantum corrections to the
expansion history of classical inflation \cite{Miao:2015oba,Liao:2018sci,
Kyriazis:2019xgj,Miao:2019bnq,Miao:2020zeh,Sivasankaran:2020dzp,Katuwal:2021kry}.
Another obvious application is for the study of phase transitions in the early 
universe \cite{Ema:2016dny}.

\centerline{\bf Acknowledgements}

This work was partially supported by Taiwan MOST grant 110-2112-M-006-026; 
by NSF grants PHY-1912484 and PHY-2207514; and by the Institute for 
Fundamental Theory at the University of Florida.

\end{document}